%% file: cortese.tex
\documentclass[float,epsfig,usenatbib]{mnras}

\usepackage{graphicx}
\usepackage{appendix}
\usepackage{url}
\usepackage{psfrag}
\usepackage{amsmath,amssymb}
\usepackage{threeparttable}
\usepackage{float}
\usepackage{subfig}
\usepackage{afterpage}
\usepackage{hyperref}

\def\apj{ApJ}
\def\apjl{ApJL}
\def\mnras{MNRAS}
\def\apss{Ap\&SS}
\def\pasp{PASP}
\def\aj{AJ}
\def\araa{ARA\&A}
\def\aap{A\&A}
\def\apjs{ApJS}

\title[The link between kinematics and optical morphology]{The SAMI Galaxy Survey: the link between angular momentum and optical morphology}
\author[L. Cortese et al.]
{L. Cortese\thanks{luca.cortese@uwa.edu.au}$^{1}$, L.~M.~R. Fogarty$^{2,3}$, K. Bekki$^{1}$, J. van de Sande$^{2}$, W. Couch$^{4}$, \newauthor 
B. Catinella$^{1}$, M. Colless$^{3,5}$, D. Obreschkow$^{1,3}$, D. Taranu$^{1,3}$, E. Tescari$^{3,6}$, 
D. Barat$^{5}$, \newauthor J. Bland-Hawthorn$^{2}$, J. Bloom$^{2}$, J.~J. Bryant$^{2,3,4}$, M. Cluver$^{7}$, S.~M. Croom$^{2,3}$, \newauthor 
M.~J. Drinkwater$^{3,8}$, F. d'Eugenio$^{5}$, I.~S. Konstantopoulos$^{4,9}$, A. Lopez-Sanchez$^{4,10}$,\newauthor S. Mahajan$^{11}$,  N. Scott$^{2,3}$, 
C. Tonini$^{6}$,
O.~I. Wong$^{1}$, J.~T. Allen$^{2,3}$, S. Brough$^{3,4}$,\newauthor M. Goodwin$^{4}$, A.~W. Green$^{4}$, I.-T. Ho$^{5,12}$, L.~S. Kelvin$^{13}$, 
J.~S. Lawrence$^{4}$, \newauthor N.~P.~F. Lorente$^{4}$, A.~M. Medling$^{5}$, M.~S. Owers$^{4,10}$, S. Richards$^{2,3,4}$,  R. Sharp$^{3,5}$, \newauthor S.~M. Sweet$^{5}$\\
$^1$International Centre for Radio Astronomy Research, The University of Western Australia, 35 Stirling Highway, Crawley WA 6009, Australia\\
$^2$Sydney Institute for Astronomy, School of Physics, The University of Sydney, Sydney NSW 2006, Australia\\
$^3$ARC Centre of Excellence for All-Sky Astrophysics (CAASTRO)\\
$^4$Australian Astronomical Observatory, PO Box 915, North Ryde NSW 1670, Australia\\
$^5$Research School of Astronomy and Astrophysics, Australian National University, Canberra ACT 2611, Australia\\
$^6$School of Physics, the University of Melbourne, Parkville, VIC 3010, Australia\\
$^7$Department of Physics and Astronomy, University of the Western Cape, Robert Sobukwe Road, Bellville, 7535, South Africa\\
$^8$School of Mathematics and Physics, University of Queensland, QLD 4072, Australia\\
$^9$Envizi Group Suite 213, National Innovation Centre, Australian Technology Park, 4 Cornwallis Street, Eveleigh NSW 2015, Australia\\
$^{10}$Department of Physics and Astronomy, Macquarie University, NSW 2109, Australia\\
$^{11}$Indian Institute of Science Education and Research Mohali-IISERM, Knowledge City, Sector 81, Manauli, P.O. 140306, India\\
$^{12}$Institute for Astronomy, University of Hawaii, 2680 Woodlawn Drive, Honolulu, HI 96822, USA\\
$^{13}$Astrophysics Research Institute, Liverpool John Moores University, IC2, Liverpool Science Park, 146 Brownlow Hill, Liverpool, L3 5RF, UK
}
\date{}

\begin{document}
\newcommand{\Zsolar}{\mbox{$\,\rm Z_{\odot}$}}
\newcommand{\Msolar}{\mbox{$\,\rm M_{\odot}$}}
\newcommand{\Lsolar}{\mbox{$\,\rm L_{\odot}$}}
\newcommand{\xs}{$\chi^{2}$}
\newcommand{\dxs}{$\Delta\chi^{2}$}
\newcommand{\xsn}{$\chi^{2}_{\nu}$}
\newcommand{\ls}{{\tiny \( \stackrel{<}{\sim}\)}}
\newcommand{\gs}{{\tiny \( \stackrel{>}{\sim}\)}}
\newcommand{\asec}{$^{\prime\prime}$}
\newcommand{\amin}{$^{\prime}$}
\newcommand{\mstar}{\mbox{$M_{*}$}}
\newcommand{\hi}{H{\sc i}}
\newcommand{\hii}{H{\sc ii}\ }
\newcommand{\kms}{km~s$^{-1}$\ }

\maketitle

\label{firstpage}

\begin{abstract}
We investigate the relationship between stellar and gas specific angular momentum $j$, stellar 
mass $M_{*}$ and optical morphology for a sample of 488 galaxies extracted from the 
SAMI Galaxy Survey. We find that $j$, measured within one effective radius, monotonically increases with $M_{*}$ and that, for 
$M_{*}>$10$^{9.5}$ M$_{\odot}$, the scatter in this relation strongly correlates with optical morphology 
(i.e., visual classification and S\'{e}rsic index). These findings confirm that massive galaxies of all types lie on a plane relating mass, 
angular momentum and stellar light distribution, and suggest that the large-scale morphology of a galaxy is 
regulated by its mass and dynamical state. We show that the significant scatter in the $M_{*}-j$ relation 
is accounted for by the fact that, at fixed stellar mass, the contribution of ordered motions to the dynamical support of galaxies 
varies by at least a factor of three. Indeed, the stellar spin parameter (quantified via $\lambda_R$) correlates strongly with S\'{e}rsic and concentration indices. 
This correlation is particularly strong once slow-rotators are removed from the sample, showing that 
late-type galaxies and early-type fast rotators form a continuous class of objects in terms of their kinematic properties.
\end{abstract}


\section{Introduction}
Since the dawn of extragalactic astronomy, it has been clear that galaxies show an impressive variety of shapes and sizes. 
Despite this diversity, astronomers soon realised that galaxies can be grouped into distinct families according 
to their visual appearance (e.g., \citealp{herschel1786,rosse1850}). Particularly successful have been the classification 
schemes proposed by \cite{reynolds1920} and \cite{hubble26}, now generally known as the Hubble sequence (see also \citealp{devauc59,vandberg76}).
After nearly a century, the Hubble sequence is still a crucial element in our theoretical framework of galaxy formation and evolution, and understanding its origin remains a challenge for current astronomical research.


Before the advent of charge-coupled devices (CCDs), galaxies were almost always classified 
via visual inspection following the Hubble classification \citep{ugc,rc3}. The high quality of photographic 
plates, combined with the proximity of the galaxies studied, allowed astronomers to notice tiny 
details in the morphology of galaxies and discriminate between various sub-classes in the Hubble sequence. 
Indeed, some of the most accurate morphological classifications to date (e.g., \citealp{vcc}) are 
still based on analysis performed on photographic plates. 

The situation changed completely with the era of CCD-based, large-area surveys. 
Firstly, as the average distances of the galaxies studied has increased remarkably, 
the fine details (e.g., dust lanes, prominence of spiral arms, faded 
disks) needed to perform accurate visual classifications are less obvious.  
Secondly, with the number of galaxies imaged increasing 
from a few thousands to millions, by-eye classification has become 
inefficient without the help of citizen science \citep{lintott08}. 
Thirdly, and perhaps most importantly, the Hubble scheme turned out not to be ideal for 
a quantitative comparison with theoretical models, as it is challenging to apply 
the same selection criteria used for observations to simulated data. 

Thus, in the last few decades, we have seen the emergence of a plethora 
of new `morphological indicators' based on the stellar distribution 
(e.g., \citealp{abraham94,bershady2000,goto03,lotz04}), optical colour of galaxies 
(e.g., \citealp{strateva01,chilingarian12}), or combinations of the two (e.g., \citealp{conselice99,banerji10,vulcani14}), 
aimed at providing a more modern view of the Hubble sequence and an easier comparison with numerical simulations. 
These morphological indicators are now common practice, and have generally replaced visual classification as a tool 
for dividing galaxies into different families.
However, despite their success and applicability to large samples of galaxies, such techniques sometimes fail to discriminate between different classes of objects. 
Particularly challenging is the regime of massive, bulge-dominated, optically-red 
galaxies where structural parameters and colours alone cannot always distinguish 
between rotationally- and dispersion-supported systems \citep{scodeggio02,emsellem07,emsellem11}, or 
between quiescent and star-forming galaxies \citep{cortese12c}. 
The main issue is that all of the above classification schemes, even when combined, 
are incomplete and are missing information about some crucial physical 
properties of galaxies such as their kinematics or star formation activity. 

In particular, it has been clear for decades that information on the stellar and gas kinematics 
can provide us with a more physically-motivated morphological classification 
(e.g., \citealp{fall83,kormendy93,kormendy04,snyder14,teklu15}). After all, the common assumption beyond the bulge vs. disk bimodality 
is that bulges are mostly supported by random motions, whereas disks are primarily 
supported by rotation. However, until very recently, the lack of resolved spectroscopic 
surveys for large, representative samples of galaxies has limited our ability to quantify 
the link between galaxy kinematics and morphology. 

Thanks to significant technical improvements, 
integral field spectroscopic (IFS) surveys of thousands of galaxies are now a reality. 
Pioneers in this new field have been the Spectrographic Areal Unit for Research on Optical Nebulae (SAURON, \citealp{bacon01}) and ATLAS$^{\rm 3D}$ \citep{cappellari11} 
surveys. By taking advantage of resolved stellar kinematics out to one effective radius ($r_{e}$), 
these projects have shown that the kinematic properties of early-type galaxies are not 
strongly correlated with their stellar light distribution \citep{krajnovic13}. They thus proposed a new 
classification scheme where early-types  
are divided into fast and slow rotators depending on the value of their spin 
\citep{emsellem11}, quantified via the $\lambda_R$ parameter \citep{emsellem07}.
Interestingly, it is still a matter of debate whether or not these conclusions hold once $\lambda_R$ 
is measured including the outer parts of galaxies \citep{foster13,arnold14}.

Two complementary approaches would naturally extend on existing kinematic studies. 
First, deeper, spatially resolved spectroscopy reaching larger galactic radii is needed to capture most 
of the angular momentum. Second, a uniform kinematic analysis of galaxies of all Hubble types is required 
to build a unified picture of the role of kinematics in galaxy evolution. 
Progress in both directions has been 
made by \citet[hereafter RF12]{romanowsky12} using a combination of stellar and gas kinematic measurements 
from the literature. They investigated the stellar mass ($M_{*}$) vs. specific angular 
momentum ($j$, the angular momentum per unit of mass) relation to quantify the connection between $j$ 
and morphology. Following the original work of \cite{fall83}, 
they showed that the scatter in the $M_{*}$-$j$ relation correlates 
with morphology (i.e., visual classification or bulge-to-total ratio) across the entire Hubble sequence. 
This suggests that, also among early-type galaxies, optical morphology statistically correlates with kinematics. 
Unfortunately, a comparison between RF12 and ATLAS$^{\rm 3D}$ is not straightforward. Not only did RF12 mainly take advantage 
of long-slit spectroscopy and not 2D resolved maps, but they also measured the total angular momentum of galaxies 
while the ATLAS$^{\rm 3D}$ work is based on the spin parameter estimated within the inner one effective radius. 

For late-type galaxies, \citet[hereafter OG14]{obreschkow14} recently improved on this limitation by taking 
advantage of resolved \hi\ velocity maps for 16 late-type galaxies from the The HI Nearby Galaxy Survey (THINGS, \citealp{things}). 
They revealed an even tighter relation between $M_{*}$, $j$ and the bulge-to-disk ratio.
However, because their sample included only late-type galaxies and $j$ is integrated 
across the entire disk, a comparison with ATLAS$^{\rm 3D}$ results is also impossible. 

To make further progress in this field, we need spatially-resolved velocity maps across 
the whole range of galaxy morphologies. The Sydney-AAO Multi-object Integral field (SAMI, \citealp{croom12}) Galaxy Survey \citep{bryant14}, 
the first large IFS survey, provides an ideal sample for which such an investigation can be carried out now.
Like all current IFS surveys, SAMI does not allow us to 
trace gas and stellar kinematics up to, or beyond, one optical radius for a statistically large number of objects. 

In this paper, we take advantage of SAMI data to extend the works of ATLAS$^{\rm 3D}$, RF12 and OG14 by investigating 
the role played by stellar and gas kinematics, within one effective radius, in shaping galaxy morphology 
across the entire Hubble sequence. 
The large number statistics, high-quality two-dimensional velocity maps and the wide range of galaxy properties 
provided by the SAMI Galaxy Survey not only allow us to ease the tension between previous works, but also provide 
us with a unique window on the physical link between stellar density distribution, spin and angular momentum. 

This paper is organized as follows. In Sec. 2 we describe the SAMI Galaxy Survey sample, the procedure to estimate 
stellar and gas velocity fields, and the ancillary data used in this paper. In Sec. 3 we investigate the 
link between central stellar and gas specific angular momentum, stellar mass, and optical morphology. 
In Sec. 4, we show the role played by the spin parameter, estimated via $\lambda_R$, in the scatter 
of the $M_{*}$-$j_{*}$ relation. In Sec. 5, we compare our results with the predictions of theoretical models. 
Finally, the implications of our results are discussed in Sec. 6.

Throughout this paper, we use a flat $\Lambda$ cold dark matter concordance 
cosmology: $H_{0}$ = 70 km s$^{-1}$ Mpc$^{-1}$, $\Omega_{0}$=0.3, $\Omega_{\Lambda}$=0.7.

\begin{figure*}
\includegraphics[width=8.5cm]{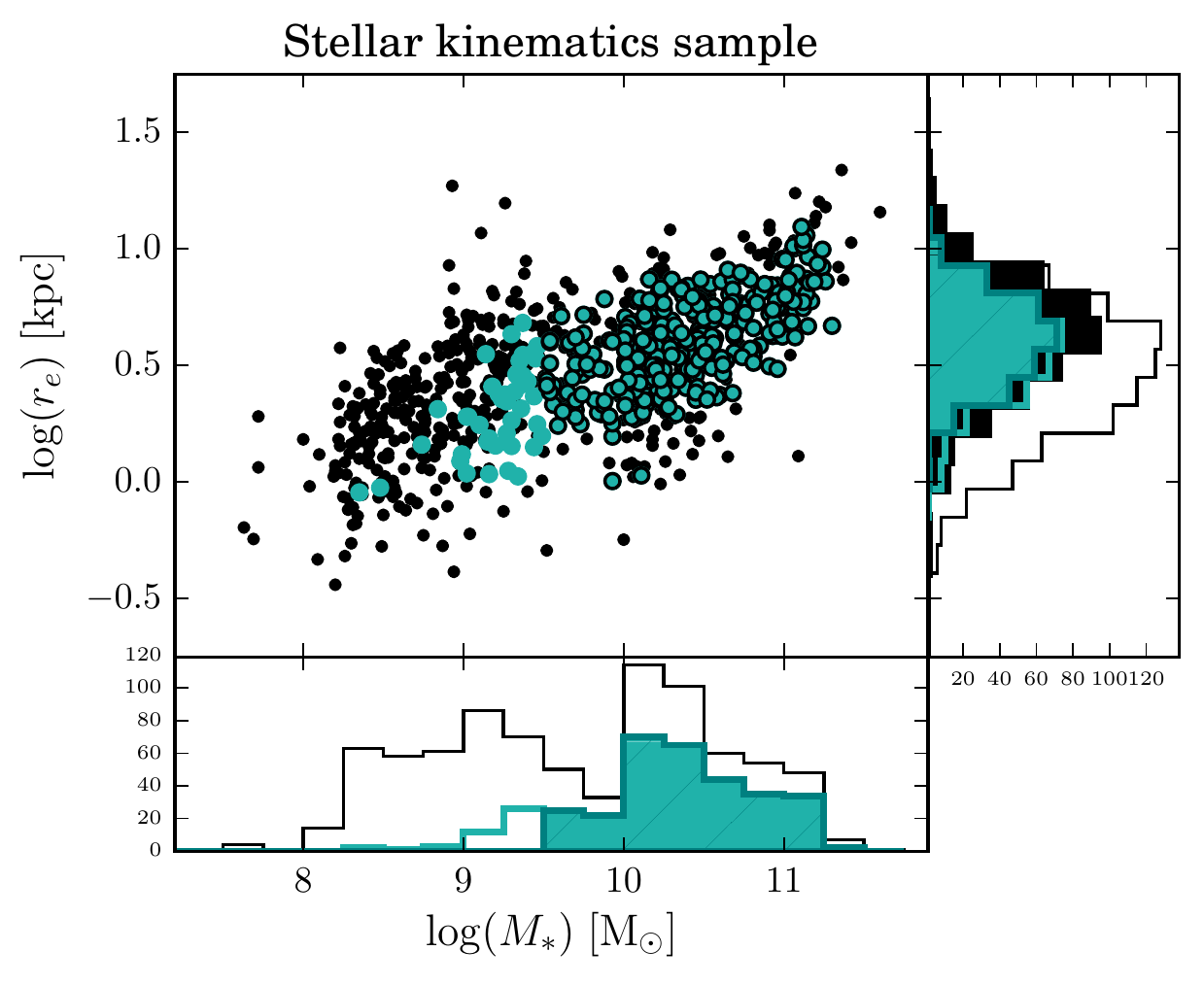}
\includegraphics[width=8.5cm]{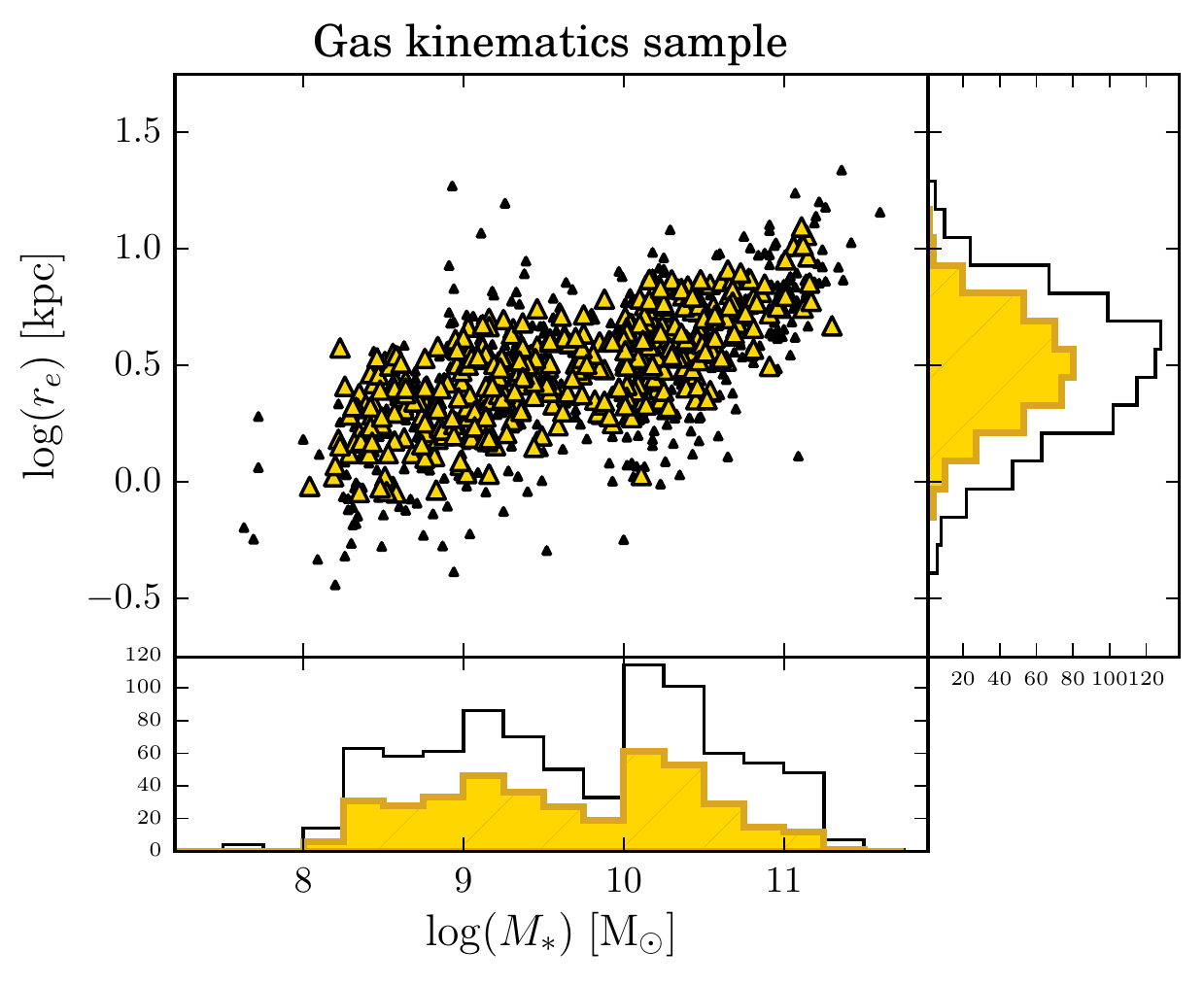}
\hspace*{-18mm}
\includegraphics[width=6.3cm]{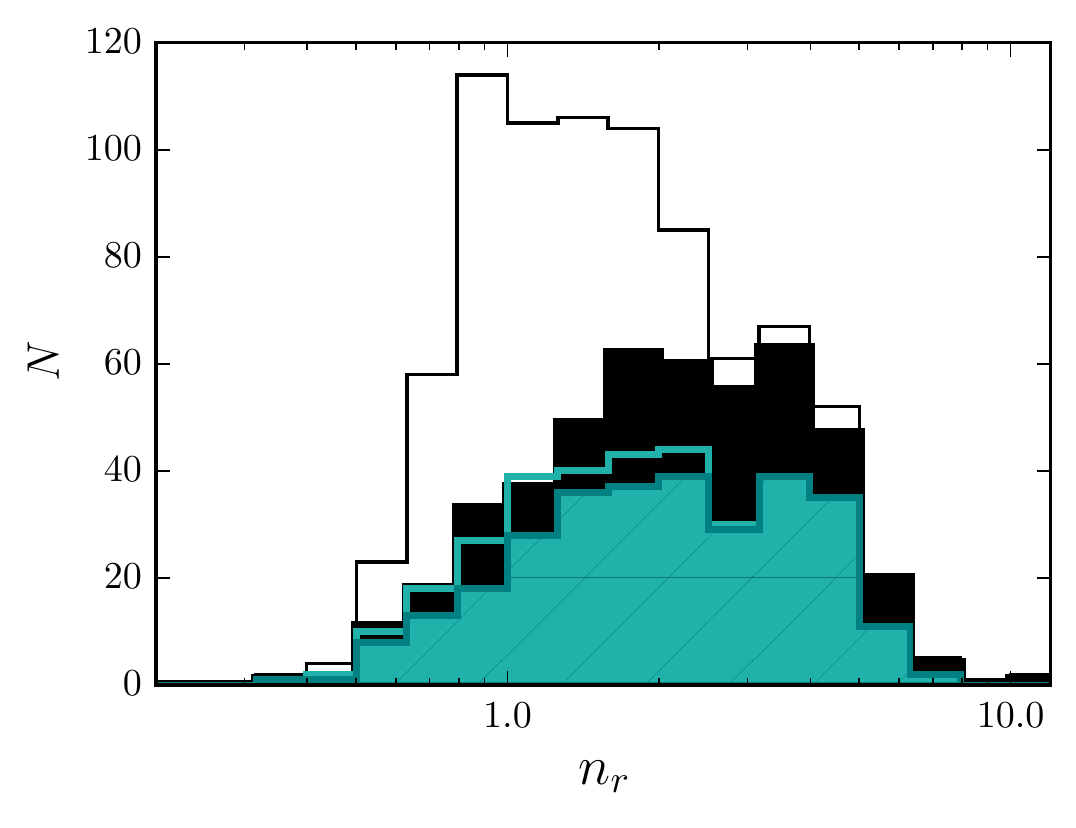}
\hskip 60pt
\includegraphics[width=6.3cm]{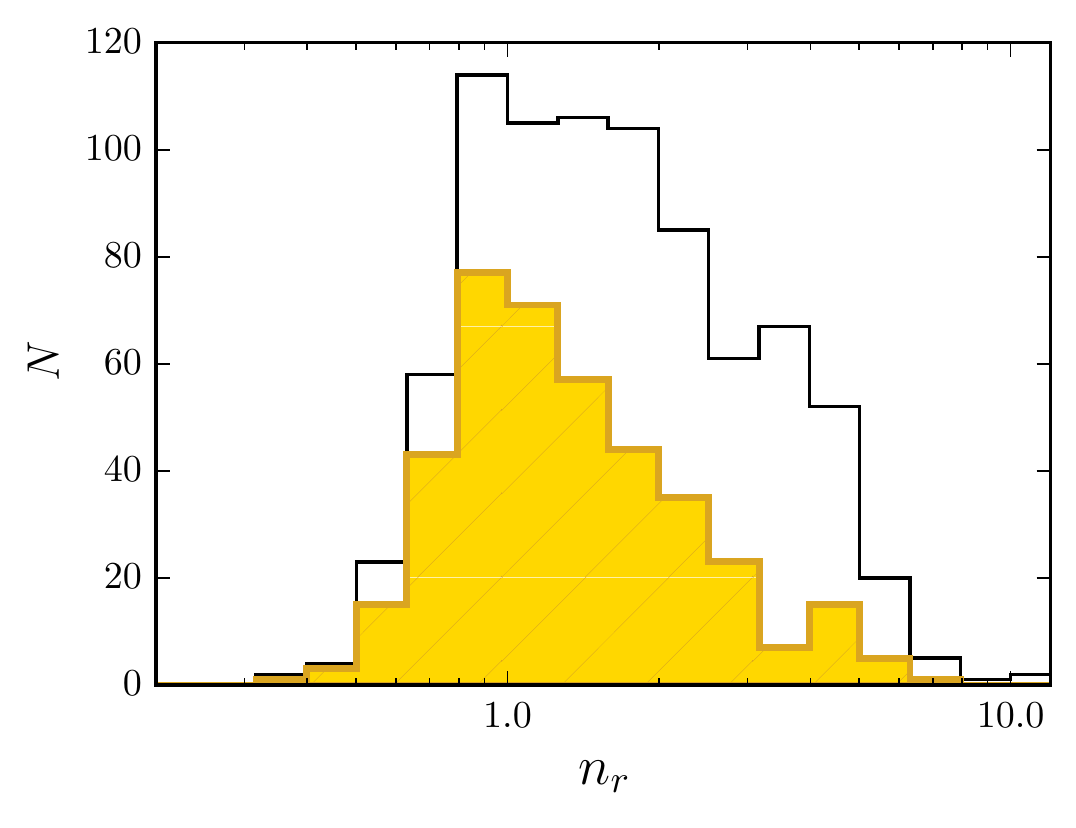}
\caption{{\it Left panels}: The $M_{*}$-$r_{e}$ relation, $M_{*}$ and $r_{e}$ distributions (top), 
and S\'{e}rsic index distribution (bottom) for galaxies with stellar kinematics in our sample. 
The black points and empty histograms show our parent sample of 824 galaxies. 
Teal points and teal empty histograms are galaxies with reliable stellar kinematics as defined in Sec.~2.1;  
black-circled teal points and teal filled histogram show our final sample after imposing a cut 
at $M_{*}$=10$^{9.5}$ M$_{\odot}$. The black filled histogram shows the 
distribution for galaxies in our parent sample with $M_{*}>$10$^{9.5}$ M$_{\odot}$. 
{\it Right panels}: Same as left for the sample with reliable gas kinematics. Note that no cut in $M_{*}$ 
has been applied.}
\label{sample}
\end{figure*}

\section{The data}
The SAMI Galaxy Survey is targeting $\sim$3400 galaxies in the redshift 
range 0.004$<z<$0.095 with the SAMI integral field unit, installed at the 3.9m Anglo-Australian Telescope. 
The main goal 
of this survey is to provide a complete census of the resolved optical 
properties of nearby galaxies (e.g., star formation rate, age, metallicity, kinematics) 
across a wide range of environments \citep{bryant14}. 

SAMI takes advantage of photonic imaging bundles (`hexabundles', \citealp{bland11,hexa14}) 
to simultaneously observe 12 galaxies across a 1 degree field of view.
Each hexabundle is composed of 61 optical fibres, each with a diameter of $\sim$1.6\arcsec, 
covering a total circular field of view of $\sim$14.7\arcsec\ in diameter. 
SAMI fibres are fed into the AAOmega dual-beam spectrograph \citep{sharp2006}, providing 
coverage of the 3700-5700 \AA\ and 6300-7400 \AA\ wavelength ranges at resolutions 
R$\sim$1730 and R$\sim$4500, respectively. 
These correspond to a velocity full-width at half-maximum of $\sim$170 \kms in the blue, 
and $\sim$65 \kms in the red.

In this paper, we focus on a subsample of the 824 galaxies from the last SAMI internal data release (v0.9 - October 2015) 
in the footprint of the Galaxy And Mass Assembly survey (GAMA, \citealp{gama}) 
for the wealth of multiwavelength data available. 
SAMI pointings targeting specific clusters outside the GAMA footprint 
(see \citealp{bryant14} and Owers et al. in prep.) are not included.
A detailed description of the data reduction 
technique is presented in \cite{sharp14} and \cite{allen14}. 
As described in \cite{bryant14}, the configuration of each SAMI plate is done 
to maximize the number of objects observable within a SAMI field of view, and no 
pre-selection on morphology or environment is introduced during the tiling process.

\subsection{Stellar and ionised gas kinematics}
Stellar and ionized gas line-of-sight velocity and velocity dispersion two-dimensional maps 
were obtained from the resampled SAMI cubes (0.5\arcsec$\times$0.5\arcsec spaxel size) 
as follows. 

Stellar line-of-sight velocity and intrinsic dispersion maps were extracted from the 
SAMI cubes by using the penalised pixel-fitting routine p\textsc{pxf}, developed by \cite{cappellari2004}.
We fitted the blue and red channels simultaneously, after having convolved the red spectra 
to the same (i.e., lower) resolution of the blue cube and interpolated on to a grid 
with the same wavelength spacing. 

We used annular binned spectra (which follow the optical ellipticity and position angle of 
the target) with signal-to-noise $\geq$25 for deriving optimal templates as opposed to obtaining an optimal template 
for each individual spaxel. Indeed, individual spaxels usually do not meet the signal-to-noise required 
to extract a reliable optimal template. 
For each annulus, we determined the best combination of the 985 stellar template spectra from the MILES 
stellar library \citep{miles} that is able to reproduce the galaxy 
spectrum. This best fit template is then used to fit every spaxel within 
that annulus having a signal-to-noise per spectral pixel greater than 3. 
We prefer annular to Voronoi bins because they allow us to follow more closely 
any radial gradients in the properties of stellar populations.
An extensive description of the SAMI stellar kinematics products will be presented 
in an upcoming paper (van de Sande et al., in prep.).

While the choice of optimal template is important for a reliable estimate of velocity dispersion, 
it has no significant effect on the line-of-sight velocity field, i.e., the critical parameter 
for the estimate of specific angular momentum. Indeed, we find the same results even if optimal templates 
calibrated for just a central 2\arcsec aperture are used to fit the entire SAMI field-of-view, as 
described in \cite{fogarty14} and \cite{cortese14b}.
Moreover, \cite{fogarty15} and van de Sande et al. (in prep.) have shown that, 
for the range of stellar velocity dispersions typical of the galaxies 
investigated in this work ($\sigma\geq$50 \kms), our technique is able to 
recover both dispersion and line-of-sight velocities, with no significant 
systematic bias.

Gas velocity maps were obtained using the 
new \textsc{lzifu} IDL fitting routine (\citealp{lzifu}; see also \citealp{ho14}). 
After subtracting the stellar continuum with p\textsc{pxf}, \textsc{lzifu} 
fits up to 11 strong optical emission lines ([O{\sc ii}]$\lambda\lambda$3726,29, H$\beta$, [O{\sc iii}]$\lambda\lambda$4959,5007, 
[O{\sc i}]$\lambda$6300, [N{\sc ii}]$\lambda\lambda$6548,83, H$\alpha$, and [S{\sc ii}]$\lambda\lambda$6716,31) as a simple Gaussian 
simultaneously using {\scshape mpfit} \citep{mpfit}, constraining all the lines to share the same velocity and dispersion. 
We use the reconstructed kinematic maps to measure gas rotation and intrinsic velocity dispersion.

Examples of SAMI stellar and gas velocity fields are presented in \citet[Fig.~3,4]{allen14}, \citet[Fig.~3,8]{allen15}, 
\citet[Fig.~6]{cecil15} and \citet[Fig.~7,A1]{ho16}.

\subsection{Ancillary data}
The SAMI data have been combined with multiwavelength observations obtained as part of the GAMA survey.
Stellar masses ($M_{*}$) are estimated from $g-i$ colours and $i$-band magnitudes following \cite{taylor11}, 
as described in \cite{bryant14}, assuming a \cite{chabrier} initial mass function and 
continuous, exponentially declining, star formation histories.
The typical random uncertainty on stellar masses is  $\sim$0.1 dex.
Effective radii, position angles and ellipticities are taken from the 2D one-component
S\'{e}rsic fits to the Sloan Digital Sky Survey \citep{york2000} $r$-band images presented in \cite{kelvin12}. 
As shown by \cite{lange15}, this dataset provides a good benchmark for the size distribution 
of local galaxies, and the radii estimated from S\'{e}rsic fits represent an improvement on earlier estimates based 
on circular apertures.

In order to investigate the link between stellar and gas kinematics and morphology, we use 
one parametric and one non parametric indicator, the S\'{e}rsic index measured in $r$-band ($n_{r}$, \citealp{kelvin12})  
and concentration index (defined as the ratio of the SDSS Petrosian 
radii containing 90\% and 50\% of the total $r$-band luminosity $R_{90}/R_{50}$), respectively. 
We use the SDSS Petrosian radii instead of those obtained from the one-component S\'{e}rsic fit in order 
to have two independent morphological indicators. Indeed, by construction, the concentration index
can be estimated analytically from the S\'{e}rsic index if the radii are derived from the one-component 
S\'{e}rsic fits.  

We also perform a visual morphological classification taking advantage of the SDSS DR9 \citep{sdssdr9} colour images. 
At least eight of us independently classified each galaxy following the scheme used by \cite{kelvin14}. 
First, galaxies are divided into late- and early-types according to their morphology, presence of spiral arms and/or signs of star formation. 
Then, early-types with just a bulge are classified as ellipticals (E) and 
early-types with disks as S0s. Similarly, late-type galaxies with only a disk component 
are Sc or later, while disk plus bulge late types are Sa-Sb. All votes are then combined and, for each galaxy, 
the type with at least 66\% of the votes is chosen. If no agreement is found, we combine adjacent votes into 
intermediate classes (E/S0, S0/Sa,Sbc) and, if the 66\% threshold is met, the galaxy is given the 
corresponding intermediate type. For those few cases (less than 5\% of our sample) for which even this second 
step fails, a new round of classifications is performed. However, this time the choice 
is limited to the two types most voted during the first iteration, and the galaxy is marked as 
unclassified if no agreement is reached. Just eight objects in our sample with either reliable gas or stellar kinematics were 
unclassified under this scheme. These galaxies will not appear in those plots in which objects are colour-coded by morphological type.

\subsection{Sample selection}
  
To obtain homogeneous and reliable estimates of the specific angular momentum within one effective radius, 
we first restrict our sample to those galaxies with an $r$-band effective diameter smaller than 
15\arcsec (the size of a SAMI bundle), and greater than 4\arcsec\ to make sure that our targets are resolved.
Then, following \cite{cortese14b}, we discard all galaxies for which more than 20\% of the spaxels 
have an uncertainty greater than 20 \kms and 50 \kms in the line-of-sight velocity of gas and stars, respectively. 
This additional cut ensures that we restrict our analysis to those galaxies for which the gas and stellar 
kinematic properties are reliable. Finally, we visually inspect each velocity map and remove problematic 
cases (e.g., contamination by foreground/background objects, disturbed systems for which the photometric ellipticity and/or position angles 
are highly inconsistent with the orientation of velocity field, etc.; $\sim$ 10\% of the remaining sample). 
After all these cuts, we are left with 397 and 341 galaxies with reliable gas and stellar kinematics, respectively.

To investigate the parameter space covered by galaxies with reliable 2D stellar or gas kinematics,  
in Fig. \ref{sample} we compare their $M_{*}$-$r_{e}$ relation and S\'{e}rsic index distribution (teal), 
with those of our parent sample of 824 galaxies (black). 
As clearly shown in the left panels of Fig.~\ref{sample}, for $M_{*}<$ 10$^{9.5}$ M$_{\odot}$ 
we do not recover the stellar kinematics for the entire range of sizes covered by our sample and 
preferentially lose systems with large radii. This selection bias roughly corresponds to a surface brightness limit 
at 1 $r_{e}$ of $\sim$23 mag arcsec$^{2}$ in $r$-band. Below this, our continuum signal-to-noise is too low 
to obtain reliable stellar kinematics. For this reason, we decided to limit our investigation 
of the stellar angular momentum to galaxies more massive than 10$^{9.5}$ M$_{\odot}$ (297 galaxies), where 
size and S\'{e}rsic index distributions for our final sample (filled teal histogram) are representative of 
the parent sample (filled black histogram).

Conversely, galaxies with reliable gas kinematics (yellow points and histograms in the right panel of Fig.~\ref{sample}) 
cover the same range of sizes and masses of our parent sample, although they clearly under-sample spheroid-dominated systems 
as highlighted by their S\'{e}rsic index distribution (golden histogram in Fig.~\ref{sample}). 
We will further discuss this bias in Sec.~3.2. 

In summary, our final sample is composed of 488 galaxies: 
397 and 297 galaxies with reliable gas and stellar kinematics, respectively (of which 206 galaxies have both stellar and gas kinematics).
It is clear that, while our samples of stellar and gas kinematic measurements are representative 
of the population of galaxies more massive than 10$^{9.5}$ M$_{\odot}$ and disk-dominated systems 
above 10$^{8}$ M$_{\odot}$, respectively, they are by no means complete. Although this does not significantly bias 
our investigation of the main driver for scatter in the $M_{*}$-$j$ relation, it could affect the value of the slope of the 
relation (see also \citealp{hyde2009}). Thus, 
as we will discuss later in the text, a grain of salt must be used in the interpretation of the slopes 
of the $M_{*}$-$j$ relations obtained as part of this work.



\section{The specific angular momentum}
In theory, following \cite{emsellem07}, the specific angular momentum of disks
can be estimated from 2D resolved line-of-sight velocity maps as:
\begin{equation}
\label{eq_jm}
\frac{J}{M} = \frac{\sum\limits_{k=1}^n {M_{k} R_{k} |V_{k}|}}{\sum\limits_{k=1}^n{M_{k}}}
\end{equation}
where $M_{k}$ is the total mass included in spaxel $k$, $R_{k}$ is its distance from the galaxy center 
in the plane of the disk (i.e., the de-projected radius), and $V_{k}$ is its rotational velocity.
In practice, SAMI data do not provide us with a distribution 
of total mass, rotational velocity and de-projected radius, but only with stellar light distribution and projected line-of-sight 
velocity and radius. Thus, a few approximations to Eq. \ref{eq_jm} are needed in order to 
estimate a proxy for the specific angular momentum from SAMI data. 

Firstly, assuming that the optical ellipticity is a good proxy for the galaxy inclination, 
the de-projected radius at each spaxel can be easily computed knowing the axis ratio and 
position angle of the galaxy.

Secondly, the spectral coverage of SAMI data does not allow us to construct 2D colour maps in the SDSS filters 
and use them to estimate the typical mass-to-light ratio in each spaxel (e.g., following standard recipes 
as in \citealp{bell03,zibetti09,taylor11}).
Thus, we simply substitute $M_{k}$ in Eq. \ref{eq_jm} with the average continuum 
flux across the entire wavelength range covered by SAMI, $F_{k}$. We further discuss the implications of this assumption in the next section, 
showing that it does not affect the main conclusions of this work. 

Thirdly, as IFS data provide information on the line-of-sight velocities, 
we need to correct for inclination in order to recover the rotational velocity of our system. 
We do so by assuming that, in each spaxel, the rotational velocity is given by: 
\begin{equation}
V_{k} = \frac{V_{k~los}}{\rm sin(\it i)\rm cos(\theta_{k})}
\end{equation}
where $V_{k~los}$ is the line-of-sight velocity, $\theta_{k}$ is the azimuthal angle in the galaxy coordinate 
frame (with zero corresponding to the direction perpendicular to the line of sight) and $i$ is the galaxy inclination. 
However, from observations we do not measure $\theta_{k}$ directly, but its projection on the plane of the sky $\phi_{k}$. 
Assuming a thin inclined disk with semi-major axis along the x direction:
\begin{equation}
\label{phi_eq} 
\tan(\phi_{k}) = \frac{y_{k}}{x_{k}} = \frac{b}{a} \tan(\theta_{k}) 
\end{equation}
where $x_{k}$ and $y_{k}$ are the x and y coordinates of spaxel $k$ with respect to the 
galaxy center, and $b$ and $a$ are the minor and major axes, respectively.
Thus \footnote{Note that, along the minor axis, cos($\theta_{k}$) is zero and our correction diverges. 
To avoid this we impose that cos($\theta_{k}$) cannot be smaller than 0.15. This effectively impacts 
only those spaxels within $<$1.1 arcsec from the minor axis of the galaxy: i.e., well within our 
spatial resolution.}, 
\begin{equation} 
\tan(\theta_{k}) = \frac{a}{b}\frac{y_{k}}{x_{k}} 
\end{equation}
Finally, inclinations are determined from the $r$-band axis ratio ($b/a$) as:
\begin{equation}
\cos(i)=\sqrt{\frac{(b/a)^{2}-q_{0}^{2}}{1-q_{0}^{2}}}
\end{equation}    
where $q_{0}$ is the intrinsic axial ratio of an edge-on galaxy.
The value of $q_{0}$ is highly uncertain and it is known to vary 
with the morphology and dynamical properties of galaxies within the range $\sim$0.1-0.65 (e.g., \citealp{gio97,weijmans14}). 
Here we use $q_{0}$=0.2 for all galaxies with a clear disk component (i.e., including S0s), and $q_{0}$=0.6 for visually 
classified ellipticals. We set the inclination to 90 degrees if $b/a<q_{0}$.
Our conclusions are not affected if we adopt a value of $q_{0}$ which varies smoothly with morphology. 
As already noted, our technique is based on the assumption that the optical axis ratio is 
a good proxy for the galaxy inclination. This is consistent with what has been done in previous works.

We remind the reader that our inclination correction is valid for disks, whereas 
for pure spheroids it systematically overestimates the effect of projection and thus 
the intrinsic angular momentum. As correcting velocity fields of pure spheroids for inclination 
is notoriously challenging even when accurate dynamical modeling can be performed (RF12;\citealp{weijmans14}), 
we do not attempt to derive an ad-hoc correction for pure elliptical galaxies. Instead, 
we perform our analysis on both projected and intrinsic (i.e., de-projected) 
specific angular momentum to show that our main results are independent of the inclination correction 
adopted. This is also because fewer than 10\% of galaxies in our sample (26 out of 297 objects) are classified 
as pure ellipticals (i.e., do not show the presence of a disk component).
An additional check on the reliability of our correction is presented in Sec.~5, where we 
compare our measurements with model predictions.

\begin{figure*}
\centering
\includegraphics[width=17.cm]{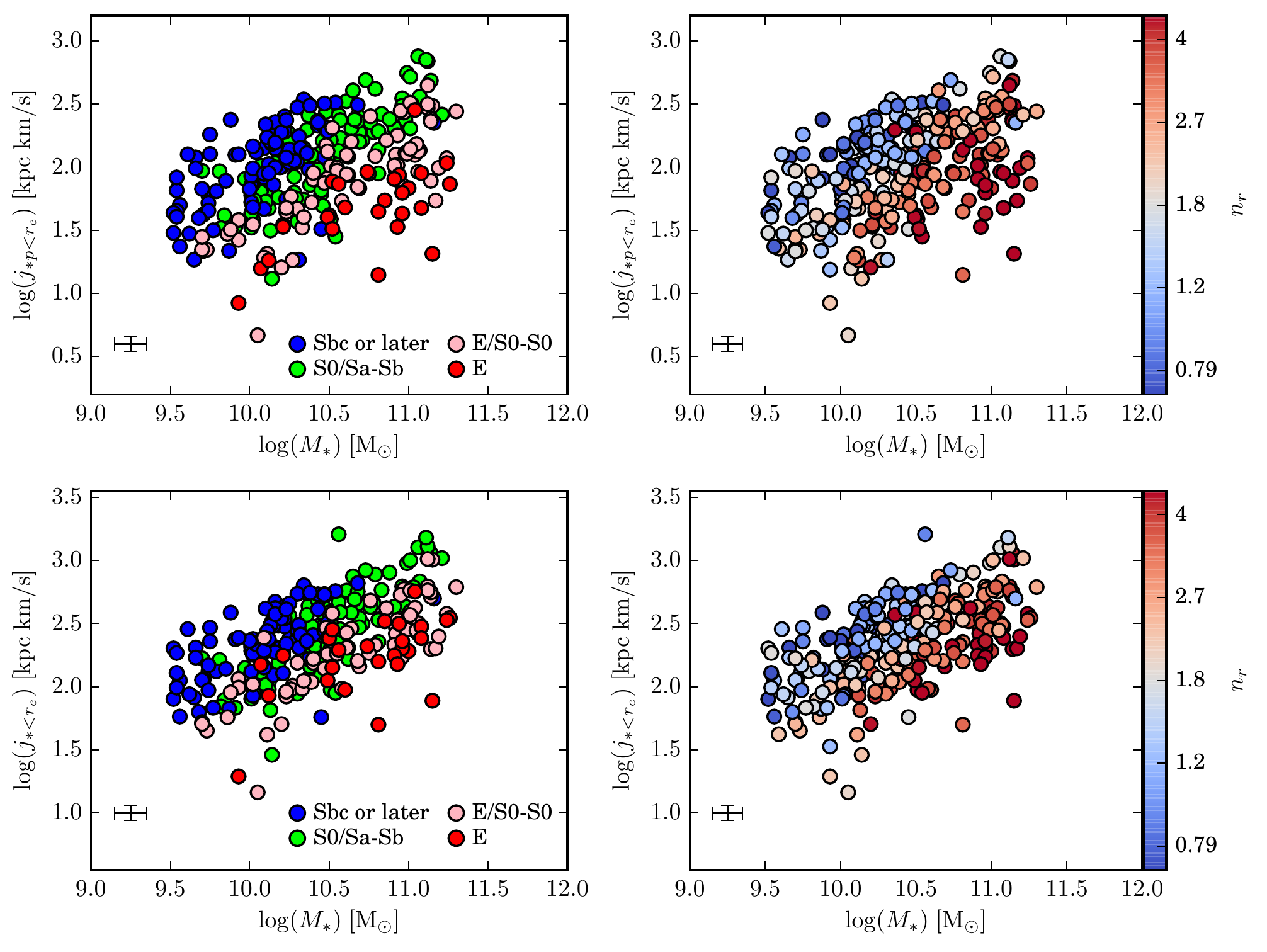}
\caption{The stellar mass versus projected ($j_{*p}$, top) and intrinsic ($j_{*}$, bottom) stellar specific angular momentum for SAMI systems. Galaxies are colour-coded by visual morphology and $r$-band S\'{e}rsic index in the left and right panels, respectively. Errorbars 
indicate mean statistical errors (i.e., not including uncertainty on inclination correction).}
\label{jm_star_morph}
\end{figure*}

To summarize, we define the projected ($j_{p}$) and intrinsic angular momentum ($j$) as:
\begin{equation}
\label{eq_jm1}
j_{p}=\frac{\sum\limits_{k=1}^n {F_{k} R_{k} |V_{k~los}|}}{\sum\limits_{k=1}^n{F_{k}}}
\end{equation}    
and
\begin{equation}
\label{eq_jm2}
j=\sum\limits_{k=1}^n  \frac{F_{k} R_{k} |V_{k~los}|}{\rm sin({\it i})cos(\theta_{k})} \times \frac{1}{\sum\limits_{k=1}^n{F_{k}}}
\end{equation}    
respectively, where here $R_{k}$ is the semi-major axis of the ellipse having minor-to-major axis ratio $b/a$ (i.e., the intrinsic galaxy radius), on which spaxel $k$ lies. 
The sum is performed including only spaxels within an ellipse of semi-major axis $R_{e}$ and 
axis ratio $b/a$. The galaxy centre is defined as the peak of the continuum emission in the SAMI cube. 
The main difference between our methodology and the one used by RF12 lies in the use of homogeneous resolved velocity maps for all galaxies 
instead of being mainly based on long-slit spectroscopy (plus the addition of multi-slit, IFS data and kinematics obtained via planetary nebulae or 
globular clusters). In our case, we can directly use the velocity information in each spaxel, following 
the 2D distribution of the velocity field. In addition, the difference in the intrinsic axial ratio 
and inclination correction imply that our $j$ are systematically lower for pure disks and higher 
for ellipticals than those obtained using RF12 technique.

Uncertainties on the specific angular momentum are estimated 
by propagating the uncertainties on continuum flux and velocity derived by p\textsc{pxf} for each spaxel, 
taking into account the covariance between individual spaxels as described in \cite{fogarty14}.
The average nominal uncertainty in $j$ is $\sim$12\%, but this does not include the effect of the 
inclination correction that likely dominates the final error.


\begin{figure*}
\centering
\includegraphics[width=15.cm]{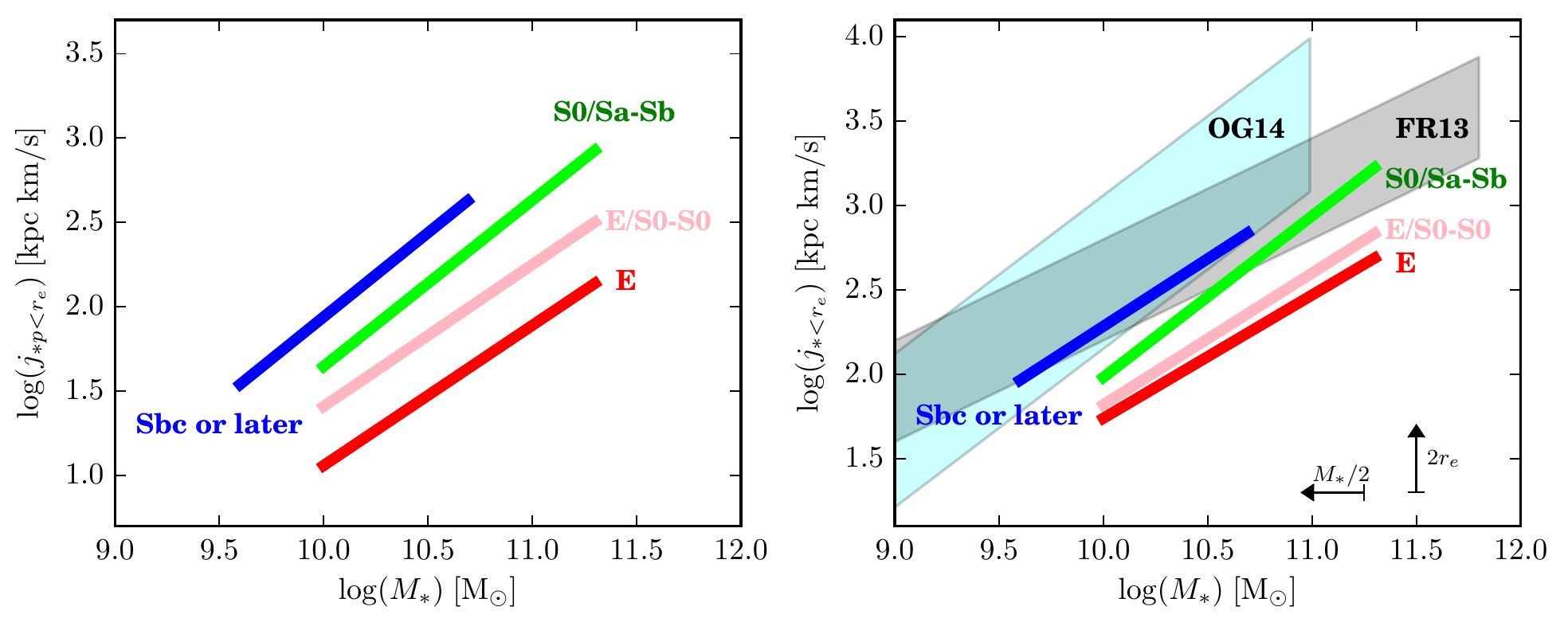}
\caption{The linear fits for the $M_{*}$-$j_{*p}$ (left) and $M_{*}$-$j_{*}$ (right) relations split 
by morphological type. The cyan and grey regions show the range covered by the relations obtained by OG14 
(including galaxies with bulge-to-total ratio between 0 and 0.3) and Fall \& Romanowsky 2013 (FR13, from disks to bulges), respectively.
We remind the reader that, for pure ellipticals, the $M_{*}$-$j_{*}$ relation must be considered as an upper limit,  
because our inclination correction is likely overestimating the projection effects in these objects. 
The vertical offset between previous studies and this work is due to the fact that we trace $j$ only out to one effective radius.
Indeed, the black arrows on the bottom-right corner of the right panel indicate how we can expect our relations to shift if 
we either plot $j_{*<r_{e}}$ as a function of half of the stellar mass of our galaxies (i.e., approximately the stellar 
mass included within one effective radius) or measure $j_{*}$ up to 2$\times r_{e}$.}
\label{jm_star_lit}
\end{figure*}

\subsection{The stellar specific angular momentum within one effective radius}
In Fig. \ref{jm_star_morph} we plot the projected and intrinsic stellar specific angular momentum, $j_{*p}$ (top panels) and $j_{*}$ (bottom panels)  
as a function of $M_{*}$ for the 297 galaxies with good stellar kinematics in our sample. Galaxies are colour-coded by visual morphology and $r$-band S\'{e}rsic 
index in the left and right panels, respectively. 
It is clear that, for the entire population, the specific angular momentum increases with stellar mass, 
and that the scatter in both relations correlates with galaxy morphology.
The scatter in the average perpendicular distance from the best-fitting bisector linear relation is $\sim$0.27 and $\sim$
0.21 dex for the projected and intrinsic case, respectively\footnote{All fits in this paper are performed by minimizing the 
orthogonal scatter while taking into account uncertainties on each variable using the hyperfit code developed by \cite{robotham15}.}.

At fixed stellar mass, disk-dominated systems have higher specific angular momentum than bulge-dominated galaxies.  
This is even clearer in Fig.~\ref{jm_star_lit}, where we present the best fits to the $M_{*}$-$j_{*}$ relation 
for the four morphological types considered here. The best-fitting parameters are presented in Table~\ref{table}. 
All four classes follow roughly parallel relations, with typical offsets of the order of 0.2-0.4 dex in $j_{*}$. 
Although the scatter in the relation is visibly reduced by the inclination correction, the effect of morphology 
in driving the spread of the intrinsic versions of the $M_{*}$-$j_{*}$ relation is still significant. 
This suggests that our findings are not an inclination effect due to the fact that, statistically, late-type galaxies are flatter than early-types. 
The only strong difference between the projected and intrinsic relations is the case of elliptical galaxies, which 
are brought closer to the relation of early-type disks once we correct for inclination. This is 
due to our conservative approach of assuming a disk geometry also for elliptical galaxies, thus likely overestimating 
the effect of projection.

In addition to inclination, it is important to investigate whether the differences shown in Fig. ~\ref{jm_star_morph} 
and ~\ref{jm_star_lit} between late- and early-type galaxies could simply be a consequence of the fact that $j_{*}$ is weighted by 
luminosity and not stellar mass. Since $j_{*}$ is a normalised quantity, it is not the absolute 
value of the mass-to-light ratio that matters\footnote{Assuming that our stellar mass estimates described 
in Sec.~2.2 properly take into account the variation of mass-to-light ratio with morphology.}, but its radial gradient. In particular, as massive late-type galaxies 
have steeper negative gradients (i.e., lower mass-to-light ratios in the outer parts) than early-type systems \citep{tortora11}, 
we could be weighting the outer parts of disks too much, thus significantly overestimating their angular momentum. 
In order to test this scenario, we estimated $j_{*}$ for our galaxies 
by assuming various mass-to-light ratio gradients. We find that even for an unrealistically large difference of 0.4 dex 
in the gradients of late- and early-type galaxies (the typical value is not greater than $\sim$0.2-0.3 dex for the stellar 
mass range of our sample; see \citealp{tortora11}), the value of $j_{*}$ changes on average by no more than 0.07 dex. This is a factor of 
four smaller than the typical difference between pure disks and late-types with bulges alone, 
and seven times smaller than the average difference between late-type disks and S0 (see also \citealp[hereafter FR13]{fall13}). 
Thus, we can definitely exclude that our trends are simply a result of an age or metallicity effect 
which directly impact the estimate of $j_{*}$. 

Lastly, as the typical seeing of the SAMI observations used in this work is of the order 
of 2.2\arcsec, beam smearing can have a non-negligible effect on the shapes of the rotation curves \citep{cecil15} and 
light distributions of our galaxies. While the decrease in velocity could lead to an underestimate of $j$, the broadening 
of the light distribution would (at least partially) balance this effect, reducing the importance of beam smearing.
Moreover, at fixed seeing, the effect of beam smearing depends on the light distribution as well as on the 
gradient of the velocity field within one effective radius. As late-type galaxies have generally larger velocity gradients 
and shallower light profiles than early-types, beam smearing could mainly artificially 
reduce (at fixed stellar mass) the difference in $j$ between disks and bulge dominated systems.
Thus, it is unlikely that our main conclusion (i.e., the role of morphology in the 
scatter of the $M_{*}$-$j_{*}$ relation) is just a consequence of beam smearing. This is also confirmed in Sec.~5, where 
we compare our observations with the predictions of (beam smearing-free) simulations.


Using numerical simulations, \cite{wu2014} found that, when random 
errors become comparable to the amplitude of the line-of-sight velocity, 
the derived angular momentum can be artificially boosted. This effect 
seems to be more prominent in slow rotators. As mentioned in Sec.~2.1, our 
tests have not shown the presence of any systematic biases in the recovered 
line-of-sight velocity. However, even if this effect is present in 
our data, it would preferentially affect slow-rotating systems, artificially 
reducing the difference between high and low angular momentum galaxies. 
Thus, the main conclusions of this work would not change.

The importance of morphology (or bulge-to-total ratio) in the scatter of the $M_{*}$-$j_{*}$ relation has 
recently been reported by RF12 and OG14. Our work confirms this finding for 
a larger sample (a factor of $\sim$3 more than RF12 and a factor of $\sim$20 more than OG14) and, 
most importantly, focuses on the effect of $j_{*}$ within one effective radius, while previous work investigated 
the total specific angular momentum. Thus, the trends shown here imply that the link between stellar kinematics 
and morphology is already well established in the inner parts of galaxies. 

This is not entirely surprising as the contribution of bulges to both the surface brightness profile and kinematical 
properties of galaxies is much more dramatic in the inner parts, which are usually well inside one effective radius. 
In Fig.~\ref{jm_star_lit} we compare our $M_{*}$-$j_{*}$ relations for different morphologies with those found by FR13 (grey area) 
and OG14 (cyan area). For FR13, the area highlighted is delimited by the $M_{*}$-$j_{*}$ relations for disks and bulges, 
while for OG14 we show the range obtained for bulge-to-total ratios varying from 0 to 0.3 (the OG14 sample does not include early type galaxies). 
The values presented in FR13 are preferred to those in RF12, as stellar mass estimates took into account 
the variation of mass-to-light ratio with morphological type.

Interestingly, the slope of our $M_{*}$-$j_{*}$ relation is intermediate between those of FR13 and OG14, 
although in general closer to the value obtained by FR13 ($\sim$0.6) than OG14 ($\sim$1). 
However, our best-fitting values should be taken with a grain of salt since, as discussed above, our sample is not 
complete. Thus, we cannot exclude the presence of a selection bias which could affect the slope of our 
relation. We stress that the most important finding here is not the slope of the relation, but the fact that its 
scatter is correlated with morphology.

What makes our results significantly different from previous work is the intercept of the relation, which is significantly 
offset towards lower specific angular momentum. As explained above, this is expected since we are tracing $j_{*}$ within one 
effective radius, thus missing the majority of the total angular momentum in galaxies, which is stored in the outer parts (RF12). 
We can test this for less than one third of our sample ($\sim$80 galaxies), for which 
we can estimate $j_{*}$ at both one and two effective radii. We find that $j_{*<2r_{e}}$ is $\sim$0.4 dex 
higher than $j_{*<r_{e}}$, making our results much more consistent with RF12 and OG14. Similarly, if we plot $j_{*}$ as a function 
of the stellar mass contained in one effective radius, our lines would shift by $\sim$0.3 dex (black arrow in the right panel of Fig.~\ref{jm_star_lit})  making them consistent with previous estimates.

Lastly, it is interesting to note that the slope of our $M_{*}$-$j_{*}$ relation, as well as the observed spread as 
a function of morphology, is in line with the recent predictions from cosmological simulations (e.g., \citealp{teklu15,genel15,pedrosa15,zavala15}).
However, all current theoretical works are focused on the total $j_{*}$, and do not investigate the $M_{*}$-$j_{*}$ relation 
within one effective radius.

\begin{figure*}
\centering
\includegraphics[width=16.5cm]{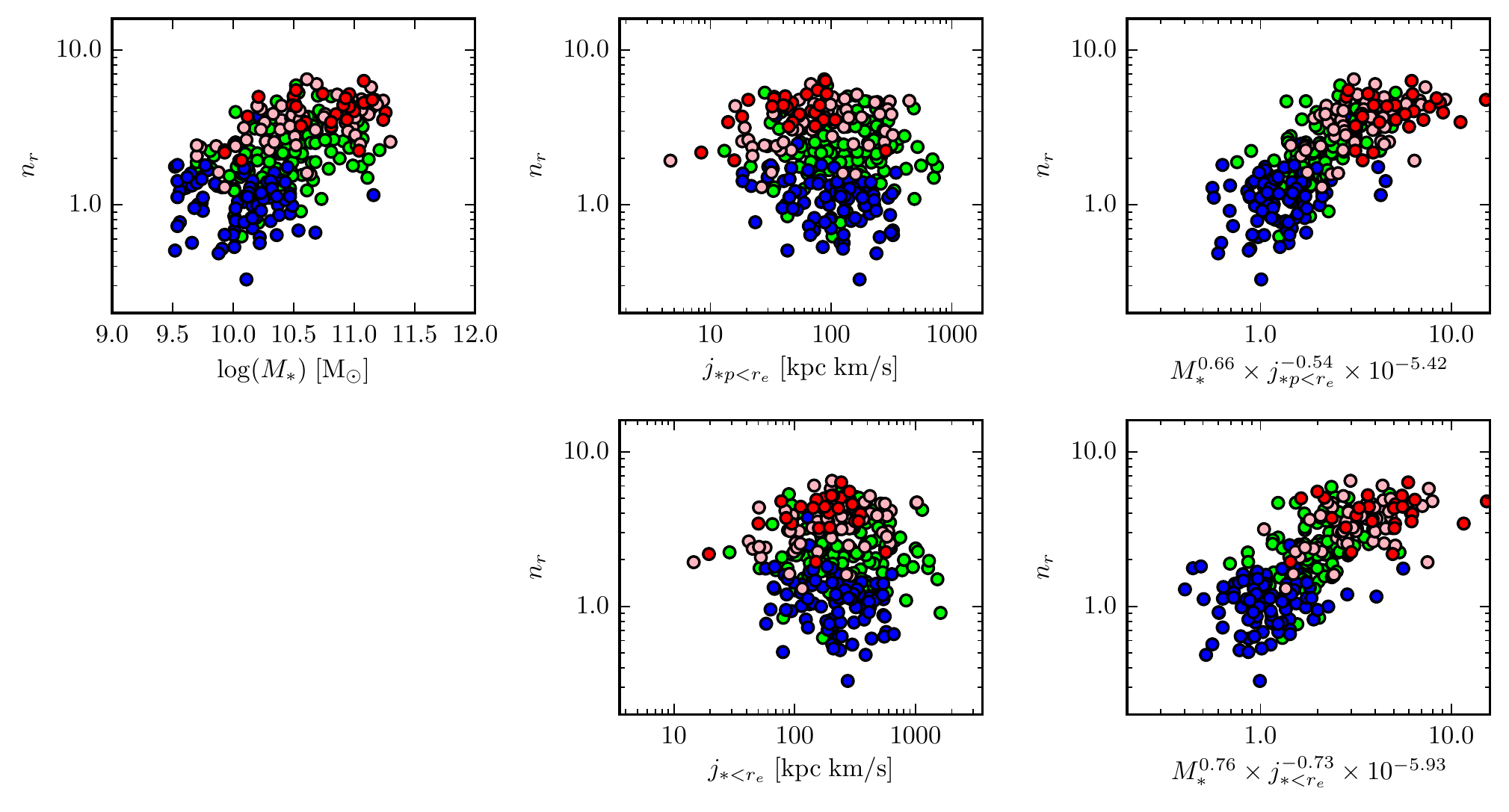}
\caption{Projections of the $M_{*}$-$j_{*}$-$n_{r}$ plane. The top row shows, from left to right, the 
$M_{*}$-$n_{r}$, $j_{*p}$-$n_{r}$, and the projection that minimises the 
scatter in $n_{r}$. The bottom row is the same as the top row, but for the intrinsic $j_{*}$. 
Symbols are color-coded by morphological type as in Fig.~\ref{jm_star_morph}.}
\label{jm_star_plane}
\end{figure*}

\input{table.tex}

\subsubsection{The central stellar specific angular momentum as a driver of morphology}
Following RF12 and OG14, the results presented in Fig.~\ref{jm_star_morph} and \ref{jm_star_lit} confirm 
that $M_{*}$, $j_{*}$ and morphology form a plane, and we 
can interpret the mix of galaxy morphologies as physically related to the spread in $j_{*}$ 
present in the local galaxy population at fixed stellar mass. The natural consequence of this result is that 
we can also look into the possibility of expressing morphological parameters such as S\'{e}rsic and concentration indices 
as a function of stellar mass and specific angular momentum. 

To do so, we fit a plane to $M_{*}$, $j_{*}$ and $n_{r}$.  
We performed this exercise on both the projected and intrinsic relations. The results are presented 
in Fig.~\ref{jm_star_plane} and compared to the $M_{*}$-$n_{r}$, and $j_{*}$-$n_{r}$ 
relations. The parameters for the best-fitting plane are presented in Table~\ref{table}.
It is clear that not only we recover the S\'{e}rsic index with $\sim$0.20 dex scatter 
(orthogonal scatter of $\sim$0.13 dex), but also that the combination of $M_{*}$ and $j_{*}$ performs significantly better than $M_{*}$ alone. 
This confirms that both the mass and the kinematic properties of galaxies play an important role in setting 
their stellar density distributions. Similar results are obtained if, instead of the S\'{e}rsic index, we 
use the SDSS-based concentration index. 
The fact that the scatter of the projected and intrinsic version of the plane are nearly the same confirms 
that our main conclusions are not an effect of inclination.
Ideally, this exercise should be performed using the bulge-to-total 
mass ratio because this is the best (i.e., more physically motivated) photometric-based morphological indicator.
Unfortunately, reliable bulge-to-disk decompositions are not yet available for our entire sample, so 
we have to postpone this analysis to future work.

Admittedly, the scatter in the $M_{*}$-$j_{*}$-$n_{r}$ plane is significantly larger than the scatter observed 
in the fundamental plane ($\sim$0.06-0.08 dex e.g., \citealp{jorgensen96,bernardi03,hyde2009,cappellari2013}) and other structural and dynamical scaling relations 
(e.g., \citealp{faber76,tully77,catinella12,cortese14b}). 
However, these relations are usually calibrated on pruned samples 
including only pre-selected morphological types (but see \citealp{cortese14b}), whereas the $M_{*}$-$j_{*}$-$n_{r}$ plane applies to 
all galaxies. Moreover, the scatter along the direction of $n$ is similar to the typical 
scatter of the main sequence of star-forming galaxies \citep{dutton2010} and of the empirical relations used to predict 
the gas content of galaxies \citep{cortese11,catinella13}.
Lastly, the orthogonal scatter of the plane is significantly smaller than that of the $M_{*}-j_{*}$ relation, 
confirming quantitatively the link between stellar mass, angular momentum and morphology.

\subsection{The gas specific angular momentum within one effective radius}
One limitation of the analysis presented above is that SAMI data allow us to trace $j_{*}$ only in galaxies with $M_{*}$\gs10$^{9.5}$ M$_{\odot}$.
Despite this, we can extend this study to lower stellar masses by measuring the kinematics 
of the ionized gas instead of the stellar component. 
Indeed, H$\alpha$ emission is detected in a significant fraction of galaxies below $M_{*}\sim$10$^{9.5}$ M$_{\odot}$ (see Fig.~\ref{sample}), 
allowing us to look at the relation between stellar mass, specific angular momentum and morphology 
across almost three orders of magnitude in stellar mass. The ionised gas specific angular momentum ($j_{gas}$) 
has been estimated from H$\alpha$ velocity maps following Eq.~\ref{eq_jm1} and \ref{eq_jm2}. 
We use the H$\alpha$ intensity map to determine $F_{k}$ in each spaxel, as we consider this a better 
proxy for the gas mass distribution in our galaxies than the optical continuum emission. However, similar results 
are obtained if the stellar emission is used instead.
\begin{figure}
\centering
\includegraphics[width=8.5cm]{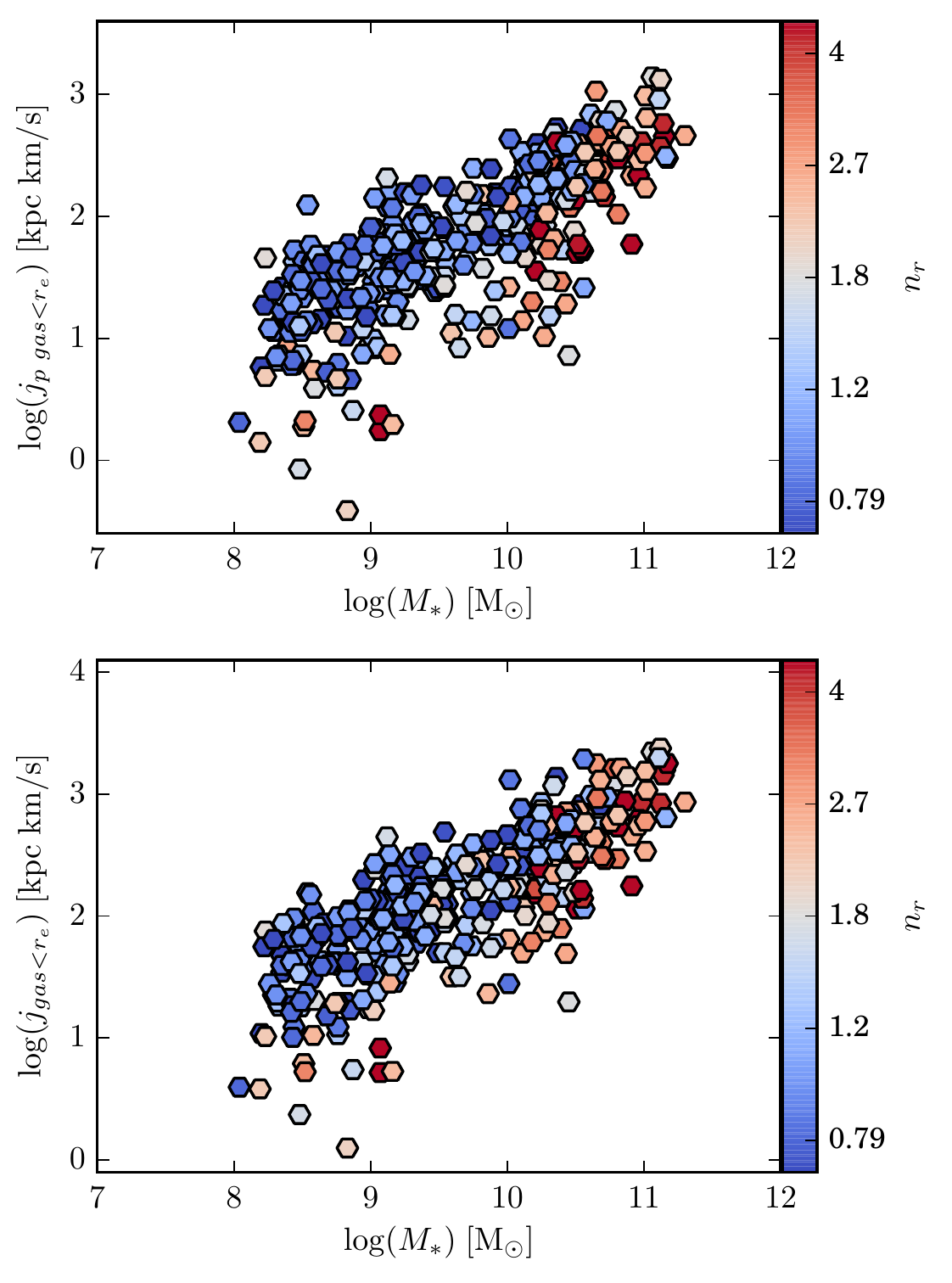}
\caption{The $M_{*}$-$j_{p~gas}$ (top)  and $M_{*}$-$j_{gas}$ (bottom) relations for SAMI galaxies. Galaxies are 
colour-coded by $r$-band S\'{e}rsic index.}
\label{jm_gas_morph}
\end{figure}

In Fig.~\ref{jm_gas_morph}, we show the $j_{p~gas}$ (top) and $j_{gas}$ (bottom) as a function of $M_{*}$. 
As in Fig.~\ref{jm_star_morph}, galaxies are colour-coded according 
to their $r$-band S\'{e}rsic index. We do not colour-code galaxies for visual morphology since, in our sample, 
nearly all galaxies below $M_{*}\sim$10$^{10}$ M$_{\odot}$ are classified as pure disks (see also 
Fig.~2 in \citealp{cortese14b}). 

The strong correlation between $M_{*}$ and $j_{gas}$ extends across the entire range of stellar 
masses covered by this work, with a scatter slightly larger than that observed for the stellar component 
(see Table ~\ref{table}). This is likely due to the use of the H$\alpha$ line emission, as its distribution is 
much more irregular than the stellar continuum. 
The slope of the $M_{*}$-$j_{gas}$ relation ($\sim$ 0.65) is consistent with the one obtained for $j_{*}$ ($\sim$0.64). 

Interestingly, the values of stellar and gas specific angular momentum (and hence 
the intercept of the $M_{*}$-$j_{*}$ relation) are significantly different, with $j_{gas}$ being systematically 
larger than $j_{*}$ (average $j_{gas}/j_{*}$ ratio $\sim$0.10 dex, obtained using those galaxies 
for which we can estimate both $j_{gas}$ and $j_{*}$). This discrepancy is consistent with (and 
a direct consequence of) the difference in 
gas and stellar rotational velocities already noticed by \cite{cortese14b} in SAMI data, and it is likely 
the effect of asymmetric drift. Thus, it is clear that, in order to properly compare galaxies of different types, 
we must compare $j$ for the same baryonic component.

Contrary to what is observed for the case of $j_{*}$, S\'{e}rsic index (or any other indicator of 
galaxy morphology) is not playing a critical role in driving the scatter of the $M_{*}$-$j_{gas}$ 
relation. Only if we focus on massive galaxies, do we recover similar trends as those shown in Figs.~\ref{jm_star_morph} and ~\ref{jm_star_plane} 
for the stellar component. Unfortunately, because we do not detect H$\alpha$ emission in many massive early-type, 
bulge-dominated galaxies (see Fig.~\ref{sample}), we are missing a crucial part of the parameter space. Moreover, while both disk and 
bulge contribute to $j_{*}$, it is likely that $j_{gas}$ mainly traces the dynamics of the disk.
Consequently, the gas angular momentum is not an ideal quantity to investigate the relation between kinematics and 
morphology in the high stellar mass regime. 

Intriguingly, there is marginal evidence for an increase in the scatter in the $M_{*}$-$j_{*}$ relation with 
decreasing stellar mass. If confirmed, this may support the findings of high 
turbulence (sometimes comparable to the rotation velocities) in the ISM of dwarf galaxies \citep{cortese14b,simons15,wheeler15}.
Moreover, the fact that the scatter in the $M_{*}$-$j_{gas}$ relation is 
similar to that of the $M_{*}$-$j_{*}$ relation (and significantly larger than the observational 
uncertainty) suggests that the dynamical state of the gas is not strongly correlated with the 
stellar light distribution in a galaxy, and that there exist other physical properties  
of galaxies responsible for the scatter of the $M_{*}$-$j_{gas}$ relation. We will investigate this issue in a future work.

\section{The spin parameter}
In a theoretical framework, the scatter of the $M_{*}$-$j_{*}$ relation should, at least partially, reflect the wide range of 
kinematic properties of dark matter halos of similar mass. Given that during the growth 
of structures, halos exert tidal torques onto each other, it is natural to expect that the degree of 
rotational support can vary across a large dynamical range. However, the exact connection between 
the angular momentum of the halo and that of the stars remains an outstanding question.

The importance of ordered motions is usually 
quantified via the spin parameter $\lambda$, which is defined as: 
\begin{equation}
\label{spin}
\lambda = \frac{J|E|^{1/2}}{GM^{5/2}}
\end{equation} 
where $J$ is the angular momentum, $E$ is the total mechanical (potential plus kinetic) energy of the system, $G$ is the 
gravitational constant and $M$ is the total mass. Thus, the scatter of the $M_{*}$-$j_{*}$ relation may correlate with $\lambda$, and $\lambda$ may somehow regulate galaxy morphology (e.g., \citealp{fall1980,dalcanton97,mo98,boissier00,zavala2008}, 
but see also \citealp{scannapieco09,romanowsky12,sales12}). Intriguingly, the width of the spin parameter distribution predicted by 
simulations is $\sim$0.22 dex (e.g., \citealp{bullok01}), very close to the scatter of our $M_{*}$-$j_{*}$ relation.

Unfortunately, estimating $\lambda$ from observations is extremely challenging.
Not only physical quantities such as total energy and total mass 
are not easily derived from observations, but also Eq.~\ref{spin} strictly applies to 
the dark matter halo, and the ratio between the spin of the halo and that 
of the baryons can easily vary during the evolutionary history of galaxies \citep{scannapieco09,sharma12,teklu15}.

\begin{figure}
\centering
\includegraphics[width=8.5cm]{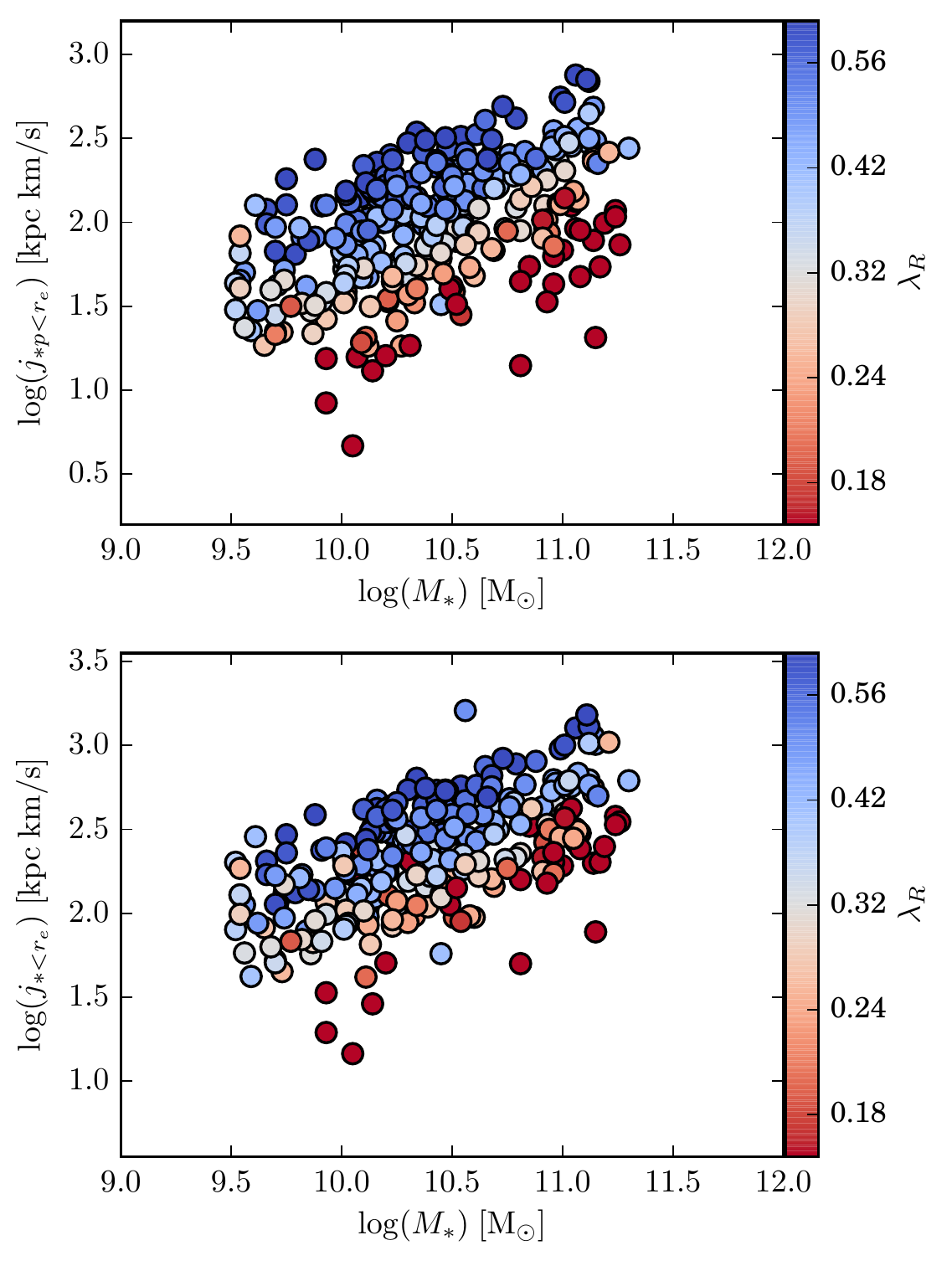}
\caption{The $M_{*}$-$j_{*p}$ (top)  and $M_{*}$-$j_{*}$ (bottom) relations with galaxies colour-coded by stellar spin parameter $\lambda_R$.}
\label{jm_lambda}
\end{figure}

\begin{figure*}
\centering
\includegraphics[width=15.cm]{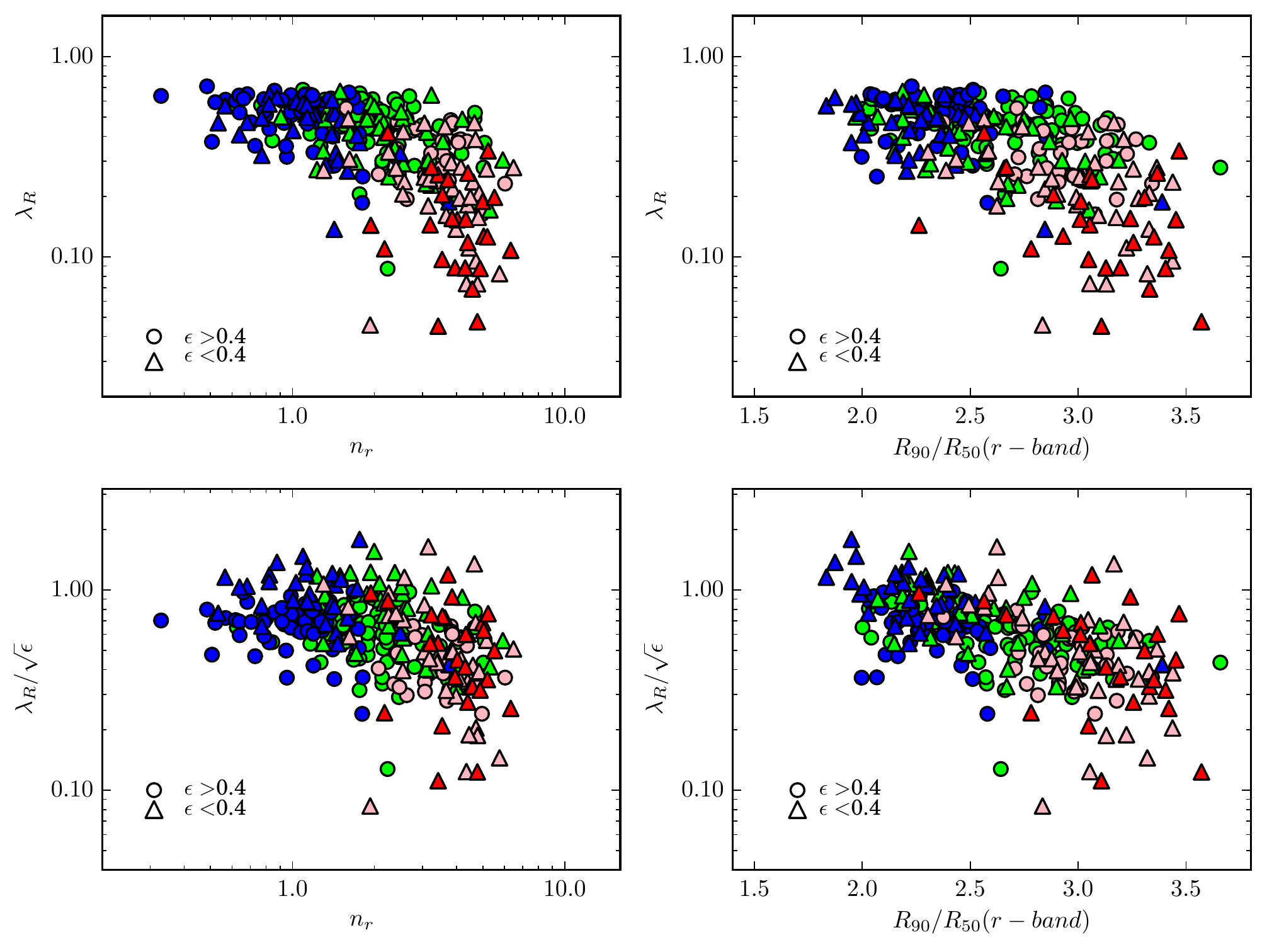}
\caption{The stellar $\lambda_R$-$n_{r}$ (top-left), $\lambda_R$-$R_{90}/R_{50}$ (top-right), $\lambda_R/\sqrt{\epsilon}$-$n_{r}$ (bottom-left), 
$\lambda_R/\sqrt{\epsilon}$-$R_{90}/R_{50}$ (bottom-right) relations for our sample. Points are colour-coded 
by visual morphology as in Fig.~\ref{jm_star_morph}. Circles and triangles indicate galaxies with ellipticities larger and 
smaller than 0.4, respectively.}
\label{lambda_morph}
\end{figure*}

In the last few years, the advent of integral field spectroscopy has made the $\lambda_{R}$ parameter \citep{emsellem07,emsellem11} 
the most commonly-used proxy for stellar spin parameter (see Appendix A in \citealp{emsellem07}): 
\begin{equation}
\label{lambdar}
\lambda_{R}=\frac{\sum\limits_{k=1}^n {F_{k} R_{k} | V_{k~los} |}}{\sum\limits_{k=1}^n{F_{k}R_{k}\sqrt{V_{k~los}^{2}+\sigma_{k}^{2}}}}
\end{equation}    
where $V_{k~los}$ and $\sigma_{k}$ are the line-of-sight and dispersion velocities in each spaxel, respectively, 
and $F_{k}$ and $R_{k}$\footnote{ We note that, although consistent with \cite{fogarty14,fogarty15}, our definition 
of $\lambda_{R}$ is different from the original definition by \cite{emsellem07}. Namely, it uses de-projected instead 
of projected radii. However, this difference does not significantly affect our findings.} are as in Eq.~\ref{eq_jm1}.
It is important to note that $\lambda_{R}$ is a projected quantity and, at face value, does not take into account 
the effect of inclination. This parameter was originally defined for 
early-type galaxies, for which inclinations are notoriously uncertain, and it has to be combined 
with the observed galaxy ellipticity to allow a proper separation between fast and slow rotators. 
Despite this possible bias, $\lambda_{R}$ is becoming commonly used for galaxies 
of all morphologies \citep{jimmy13,pracy13,tapia14,falcon15,fogarty15}, thus it is interesting to see 
how the results presented above can be interpreted in the context of this parameter. 

Fig.~\ref{jm_lambda} shows the $M_{*}$-$j_{*}$ relation, this time colour-coded by values 
of $\lambda_R$. As expected, since $j_{*}$ and $\lambda_R$ are not independent quantities,
we find that the scatter in the relation correlates strongly with 
$\lambda_R$. Indeed, the scatter in the best-fitting $M_{*}$-$j_{*}$-$\lambda_R$ plane 
is $\sim$0.08 dex, significantly smaller than in the case of the S\'{e}rsic index (see Table~\ref{table}). 
Remarkably, the best-fitting coefficients are very close to $j\sim M_{*}^{2/3}\times\lambda_{R}$, which can 
be recovered analytically (see e.g., RF12 and OG14) within the general theoretical framework of \cite{mo98}, 
assuming that $\lambda_{R}$ is proportional to halo spin parameter, and a constant halo-to-stellar mass ratio. 
The projected version of the plane has a scatter significantly smaller than the intrinsic one. 
This is simply because $\lambda_R$ is a projected quantity and thus it correlates more tightly with the scatter 
of the $M_{*}$-$j_{p*}$ relation.

From an observational point of view, since the slope of the $M_{*}$-$j_{*}$-$\lambda_R$ plane in 
the $\lambda_{R}$ projection is very close to 1, the tight $M_{*}$-$j_{*}$-$\lambda_R$ plane becomes akin\footnote{ Indeed, 
if  $j_{*}\propto \lambda_{R}M_{*}^{a}$ by simply dividing $j_{*}$ and $\lambda_{R}$, the equation for 
the plane can be re-expressed as $M_{*}\propto$ 
\bigg($\frac{\sum\limits_{k=1}^n { F_{k}R_{k}\sqrt{V_{k~los}^{2}+\sigma_{k}^{2}}} }{\sum\limits_{k=1}^n{F_{k}}}\bigg)^{1/a}$.} 
to the known relation between $M_{*}$ and $\sqrt{V^{2}+\sigma^{2}}$, \citep{cortese14b},  
which shows similar scatter ($\sim$0.1 dex) and represents a promising unified dynamical scaling relation valid for galaxies of all 
regular morphological types.

The role played by $\lambda_R$ in the scatter of the $M_{*}$-$j_{*}$ relation, combined with the results
of Sec.~3, implies that $\lambda_R$ should correlate with indicators of optical morphology such 
as $n_{r}$ and concentration index. This correlation is investigated in the top panel of Fig.~\ref{lambda_morph}. Although 
$\lambda_R$ clearly correlates with both quantities (Spearman correlation coefficient $\sim$$-$0.6), 
the relations show quite a large amount of scatter, as recently highlighted by \cite{fogarty15} 
using a smaller sample of cluster galaxies from the SAMI pilot survey (see also \citealp{falcon15}). 
This is particularly true for high S\'{e}rsic and concentration indices, where there is almost 
no correlation between $\lambda_R$ and optical morphology. 
Interestingly, this is the typical parameter space occupied by the population of `slow-rotators' 
investigated by the ATLAS$^{\rm 3D}$ survey \citep{cappellari11,emsellem11}, for which it has been claimed that optical morphology 
does not represent a good proxy for their kinematic properties \citep{krajnovic13}.

However, part of the scatter and non linearity in the $\lambda_R$-$n_{r}$ (left) 
and $\lambda_R$-$R_{90}/R_{50}$ relations is likely just a consequence of the fact that $\lambda_R$ is a projected quantity. 
As shown in Fig.~\ref{lambda_morph}, the vast majority of the outliers from the main relation are 
galaxies with ellipticities smaller than 0.4 (triangles in Fig.~\ref{lambda_morph}). 
Moreover, if we try to account for the effect of inclination by simply plotting $\lambda_R/\sqrt{\epsilon}$ instead of $\lambda_R$, 
the correlation becomes more linear, in particular for the concentration index.
Of course, this is a crude way to correct for inclination and to properly quantify projection effects, 
something outside the scope of this paper, we do require detailed dynamical modeling. Indeed, not only the inclination but also 
the anisotropy of the velocity field are needed to correct both line-of-sight and dispersion velocities. 

Thus, at this stage, we can at least safely conclude that, excluding slow-rotators, there is a good correlation between 
optical morphology and $\lambda_R$, with the value of the spin parameter decreasing with the increase of stellar 
concentration. 


\section{Comparison with models}
\label{secmodel}
The most natural interpretation of our results is that the stellar density distribution in galaxies, 
and thus their morphology, is a direct manifestation of the contribution of ordered 
motions to the dynamical support of the system. The larger the contribution of dispersion, the more 
centrally concentrated the stars are and the more closely the galaxy resembles to a bulge-dominated system. 
This is consistent with previous works that found a correlation 
between the $V/\sigma$ ratio and galaxy morphology \citep{courteau2007,catinella12}. 

However, it is important to make sure that such a scenario is not only qualitatively, but also quantitatively 
consistent with our findings. Thus, in this section we compare our results with the predictions of the 
theoretical model developed by \cite{bekki13c} for isolated galaxies. A detailed comparison with numerical simulations 
in a cosmological context will be the focus of a future paper. 
The immediate advantage of using the \cite{bekki13c} code is that, thanks to its high resolution 
($3 \times 10^5 {\rm M}_{\odot}$ in mass and 193 pc in size), we can analyse 
the output of the simulation using the same tools used for the SAMI data, and extract physical quantities 
in a consistent way. The main goal of this exercise is simply to test if the scatter in the 
$M_{*}$-$j_{*}$ relation and the relation between stellar concentration and $\lambda_R$ can be reproduced 
by increasing the mass of a fully dispersion supported bulge component. 

We use the realisations of disk galaxies presented in \cite{bekki14}. 
Briefly, a disk galaxy is assumed to consist of a dark matter halo,
a stellar and gas disk, and a stellar bulge. The gas-to-stellar mass ratio and 
the total stellar-to-dark matter disk mass ratio are set to be 0.1 and 0.06,
respectively.

\begin{figure*}
\centering
\includegraphics[width=15.cm]{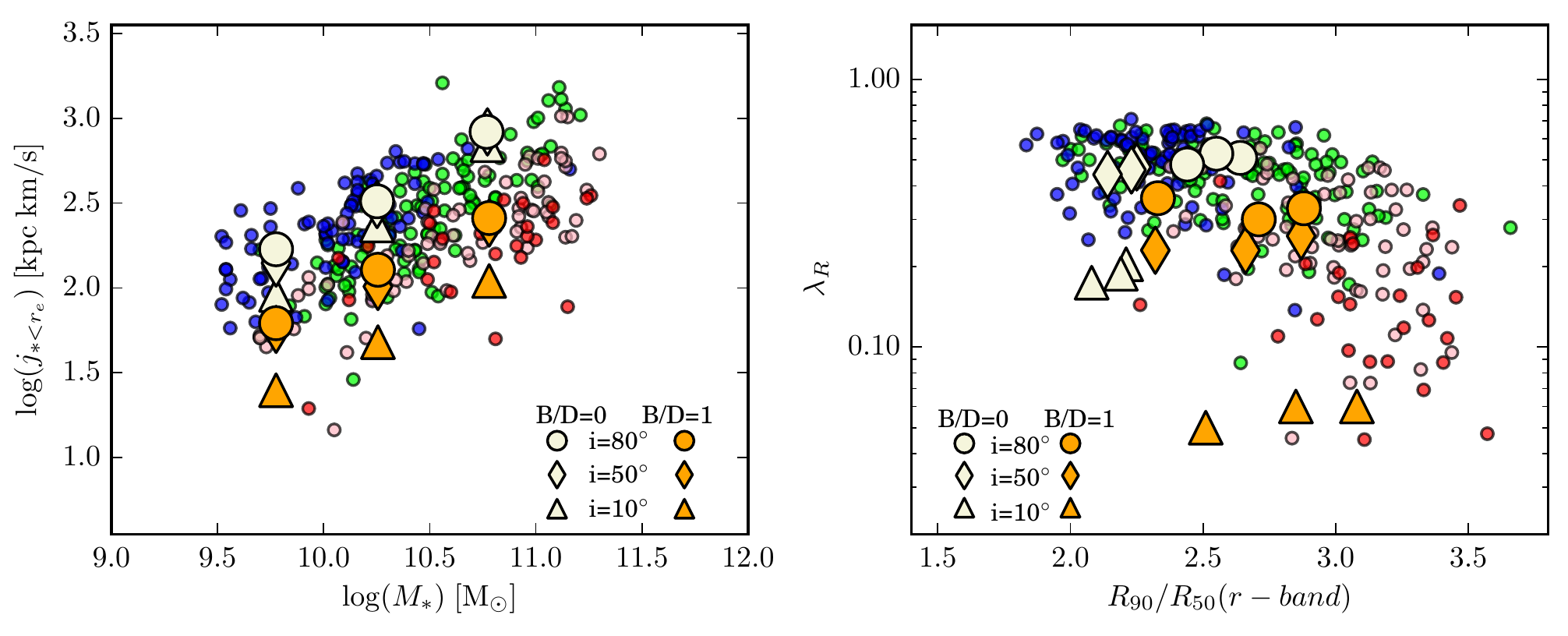}
\caption{Comparison between the observed $M_{*}$-$j_{*}$ (left) and $\lambda_R$-$R_{90}/R_{50}$ (right) relations (small circles) 
and the predictions of the model by Bekki (2013, 2014; large symbols). Beige and orange symbols indicate models for pure disks 
and disk galaxies with B/D ratio of 1, respectively. For each model, large circles, diamonds and triangles show 
inclinations of 80, 50 and 10 degrees, respectively. SAMI galaxies are colour-coded by their visual morphology as in Fig.~\ref{jm_star_morph}.}
\label{models}
\end{figure*}

The initial density profile of the dark matter halo 
is assumed to be a Navarro-Frenk-White (NFW; \citealp{nfw}) profile with concentration set to 10. The bulge component 
has a \cite{herquist90} density profile evaluated up to five scale lengths, with 
the bulge scale length set to 0.35 times that of the disk.
The bulge is assumed to be non rotating and to have an isotropic velocity dispersion. The radial 
velocity dispersion is given according to the Jeans equation for a spherical system, including 
also the mass contribution from the halo.

The radial ($R$) and vertical ($Z$) density profiles of the stellar disk are
assumed to be proportional to $\exp (-R/R_{0}) $ extending up to 5 scale lengths, and 
to ${\rm sech}^2 (Z/Z_{0})$ with scale length $Z_{0} = 0.2R_{0}$, respectively.
The size of the gas disk is twice that of the stellar one.

In addition to the rotational velocity caused by the gravitational field of the disk,
bulge, and dark halo components, the initial radial and azimuthal
velocity dispersions are assigned to the disk component according to
the epicyclic theory with Toomre's parameter $Q$ \citep{toomre64}. Here we choose 
$Q$=3.0, which appears to best match the observed stellar velocity dispersion of 
SAMI galaxies. 

We consider models for two values of bulge-to-disk stellar mass ratio ($B/D$=0 and 1)  
and three total dark matter masses (1, 0.3 and 0.1 $\times$ 10$^{12}$ M$_{\odot}$, corresponding 
to disk scale lengths of 3.5, 1.9 and 1.1 kpc). 
For each model we extract line-of-sight velocity, velocity dispersion and stellar density maps using a mesh size 
of 0.5 kpc, roughly consistent with the typical size of SAMI spaxels, and assuming 
three inclinations: 10, 50 and 80 degrees. `Observed' total stellar masses, effective radii, ellipticities 
and concentration indices\footnote{In order to be consistent with observations, effective radii are 
obtained by fitting isophotal ellipses, while the concentration index comes from the radii obtained from 
fitting circular apertures to the simulated data.} are estimated directly from the stellar mass distribution maps by fitting isophotal 
ellipses, following the technique described in \cite{hrsgalex}. 
Specific angular momentum and $\lambda_R$ are then extracted within one effective radius following Eq.~\ref{eq_jm2} and \ref{lambdar}, 
and using exactly the same software adopted for the SAMI data. The only difference between simulated and observed parameters is that 
simulated $j_{*}$ and $\lambda_R$ are weighted by stellar mass instead of luminosity.  

In Fig.~\ref{models}, we compare the model predictions with the SAMI data on the $M_{*}$-$j_{*}$ (left) and the 
$\lambda_R$-$R_{90}/R_{50}$ (right) relations. The beige and orange points show the $B/D$=0 and $B/D$=1 cases, respectively. 
The agreement between simulated and real data is encouraging. Although the slope of the $M_{*}$-$j_{*}$ relation is a direct 
consequence of the modeling (i.e., of the assumptions made on the mass and velocity profiles), it is interesting to see that we are able to quantitatively reproduce 
the difference between the pure-disk and bulge plus disk models not only in the $M_{*}$-$j_{*}$ relation, but also 
in the $\lambda_R$-$R_{90}/R_{50}$ plot. 
The difference between models with different inclinations (circles, diamonds and triangles in Fig.~\ref{models}) also gives an idea 
of the systematic uncertainty in our inclination correction. In particular, it is not surprising that the inclination correction is much 
more erroneous for bulge dominated spirals than for pure disks. For inclined early-type disks, the best-fitting 
ellipses to the entire galaxy have generally a smaller ellipticity than the disk alone, significantly affecting 
the accuracy of our inclination correction. Although Fig.~\ref{models} indicates that our inclination correction 
might underestimate $j_{*}$ for face-on bulge-dominated galaxies, we note that just a couple of galaxies in our S0 sample are face-on, whereas 
the vast majority of our targets have inclinations greater than 40 degrees where our correction appears to work properly.
 
In summary, it is clear that the presence of prominent, fully dispersion supported bulges can 
quantitatively reproduce the offset between late- and early-type galaxies in the $M_{*}$-$j_{*}$ relation 
and the trend seen between spin and concentration (see also RF12). Thus, this comparison confirms that the results presented 
in this paper can be interpreted as a simple manifestation of the physical link between the stellar density distribution 
and kinematics in galaxies across the Hubble sequence. The next step is therefore to compare our findings 
to the predictions of cosmological simulations to follow the growth of mass and angular momentum 
in galaxies in a self-consistent fashion (e.g., \citealp{snyder14,genel15,teklu15}).

\section{Discussion \& Conclusions}
The analysis presented in this paper provides quantitative evidence that 
both the stellar and gas specific angular momentum of galaxies measured within one effective 
radius strongly correlate with stellar mass. The slope of the relation across the whole sample ($\sim$0.64) is remarkably close 
to the value expected from analytical models (2/3). However, given that our sample is not complete, 
future confirmation for the exact slope of the $M_{*}$-$j_{*}$ relation is needed.

We show that for stellar masses \gs 10$^{9.5}$ M$_{\odot}$, the scatter in the $M_{*}$-$j_{*}$ relation 
is related to the stellar light distribution, hence morphology, of galaxies.
Compared to previous works, not only do we take advantage of significantly larger number 
statistics, but also, thanks to SAMI integral field spectroscopy, we are able to quantify the specific angular momentum 
using exactly the same technique for all galaxies in our sample. 

One of the most important implications of our findings is that, from a statistical point of view, 
we can quantify galaxy morphology via the kinematic properties of galaxies: 
once we know the stellar mass and specific angular momentum of an object, we 
can predict what its stellar light profile will be. In other words, galaxies lie 
on a tight plane defined by their S\'{e}rsic index, stellar mass, and specific angular momentum (see also OG14). 
A similar conclusion is reached if $j_{gas}$ is used instead of $j_{*}$. 
However, as the presence of H$\alpha$ emitting gas up to one effective radius is not widespread in early-type systems, 
it is much more challenging to use $j_{gas}$ to calibrate the $M_{*}$-$j$-$n_{r}$ plane. 

We show that, from a physical point of view, the scatter in the $M_{*}$-$j_{*}$ relation is simply a consequence 
of the fact that, at fixed stellar mass, the contribution of ordered motions to the dynamical 
support of galaxies varies by at least a factor of three. Indeed, the stellar spin parameter 
$\lambda_R$ is even more correlated with the scatter in the $M_{*}$-$j_{*}$ relation. 
This is quite remarkable considering that $\lambda_R$ is a projected quantity, not 
corrected for the effect of inclination. Intriguingly, we find that the correlation between $\lambda_R$ and 
morphology seems to break down for bulge-dominated, slow-rotator galaxies, suggesting that at fixed 
stellar concentration we can have a wide range of spin parameters. 
However, this could simply be an inclination effect, and further analysis (including accurate inclination corrections) 
are needed to determine whether or not the stellar density distribution alone is 
sufficient to isolate slow rotators \citep{emsellem11,krajnovic13}.  

Conversely, the tight relation between spin and S\'{e}rsic index observed for 
the rest of our sample shows that, when we look at their stellar distribution and kinematics, early-type fast rotators and 
late-type galaxies are not two separate classes of objects, but represent a `continuum' connecting pure-disks to 
bulge-dominated systems. Given that galaxies with a disk/rotationally-supported component are by far the most common in the local 
Universe \citep{emsellem11,kelvin14}, the ability to link their morphological properties 
to their kinematics is of critical importance for understanding the origin of the Hubble sequence. 
For example, as also illustrated by \cite{fogarty15}, our results imply that, if galaxies are really 
morphologically transformed, their stellar kinematics should be affected as well. 
Similarly, if galaxy transformation is simply a result of the quenching of the star formation 
(and of the consequent fading of the spiral arms), at fixed stellar mass passive galaxies should show the same 
kinematical properties as star-forming disks. Thus, information on the stellar kinematics of galaxies can 
allow us to investigate these scenarios, moving beyond relations such as the morphology-density relation.
Particularly promising is the ability to investigate the effect of the environment in terms of mass, star formation and angular momentum without the need 
to split galaxies by morphology, as is currently done even for the so-called kinematic morphology-density 
relation \citep{cappellari11b}.     

Although our findings are consistent with theoretical expectations, and with previous observations 
by RF12 and OG14, they may appear in contradiction with \cite{krajnovic13}, who did not find 
a correlation between S\'{e}rsic index and $\lambda_R$ for early-type galaxies in ATLAS$^{\rm 3D}$. 
However, it is easy to show that this is simply due to the fact that the ATLAS$^{\rm 3D}$ sample included 
only early-type galaxies, thus missing the large family of rotationally supported systems with 
little or no bulge component. Indeed, if we focus on early-type galaxies only, it is clear from 
the top row of Fig.~\ref{lambda_morph} that the trend disappears also in our sample, consistently with Fig.~4 in \cite{krajnovic13}. This simply supports 
the argument that early-type fast rotators and late-type galaxies should be treated as a single population, 
and it is fully consistent with the proposed revision of the Hubble tuning-fork, where 
S0s are directly linked to late-type disk galaxies and they are no longer a transition class between 
spirals and ellipticals \citep{spitzer51,vandberg76,cappellari11b,kormendy12}.

In the future, it will be important to extend our results by replacing S\'{e}rsic 
and concentration indices with accurate estimates of bulge-to-disk ratios. As bulge-to-disk 
decomposition is arguably the most physically motivated imaging-based morphological classification, 
we should find that the increase in the importance of random motions across the 
Hubble sequence is directly related to the increase of the mass in the `photometric' bulge. 

Finally, it is important to highlight the limitations of our current analysis in order 
to avoid dangerous extrapolation of our findings. Firstly, as SAMI data allow 
us mainly to investigate the inner parts of galaxies, it is possible (and perhaps 
even expected) that some of our conclusions change once the total (i.e., integrated 
up to several effective radii) angular momentum is taken into account \citep{arnold14}. 
However, the fact that RF12 reaches similar conclusions by investigating the total 
specific angular momentum is encouraging.
Secondly, due to our limited spatial resolution, we struggle to trace with extreme 
detail gas and stellar kinematics in the inner 1-2 kpc of our targets. Thus, at this stage, 
our velocity maps do not allow us to discriminate between the presence of a classical or a 
pseudo-bulge \citep{kormendy04} and determine their role in the scatter of the 
$M_{*}-j_{*}$ relation. This means that velocity maps with kiloparsec or sub-kiloparsec resolution, 
extending up to the outer edges of galaxies, will be critical for further unveiling 
the complex connection between galaxy structure and kinematics.

Nevertheless, our work already demonstrates how homogeneous estimates of the 
stellar and gas angular momentum across all galaxy types allow us 
to move beyond visual morphology and shed light on the physical origin of the Hubble sequence.

\section*{Acknowledgments}
We thank the referee for a very detailed and constructive report which significantly 
improved the quality of this manuscript.

LC acknowledges financial support from the Australian Research Council (DP130100664, DP150101734).
BC is the recipient of an Australian Research Council Future Fellowship (FT120100660).
JTA acknowledges the award of a John Stocker Postdoctoral Fellowship from the Science and Industry Endowment Fund (Australia).
MSO acknowledges the funding support from the Australian Research Council through a Future Fellowship Fellowship (FT140100255).
SB acknowledges the funding support from the Australian Research Council through a Future Fellowship (FT140101166).
AMM acknowledges the support of the Australian Research Council through Discovery project DP130103925.

Part of this work was performed on the gSTAR national facility at Swinburne University of Technology. gSTAR is funded by Swinburne and the Australian Government’s Education Investment Fund.

The SAMI Galaxy Survey is based on observations made at the Anglo-Australian Telescope. The Sydney-AAO Multi-object Integral field spectrograph (SAMI) was developed jointly by the University of Sydney and the Australian Astronomical Observatory. The SAMI input catalogue is based on data taken from the Sloan Digital Sky Survey, the GAMA Survey and the VST ATLAS Survey. The SAMI Galaxy Survey is funded by the Australian Research Council Centre of Excellence for All-sky Astrophysics (CAASTRO), through project number CE110001020, and other participating institutions. The SAMI Galaxy Survey website is \url{http://sami-survey.org/}.

GAMA is a joint European-Australasian project based around a spectroscopic campaign using the Anglo-Australian Telescope. The GAMA input catalogue is based on data taken from the Sloan Digital Sky Survey and the UKIRT Infrared Deep Sky Survey. Complementary imaging of the GAMA regions is being obtained by a number of independent survey programs including GALEX MIS, VST KiDS, VISTA VIKING, WISE, Herschel-ATLAS, GMRT and ASKAP providing UV to radio coverage. GAMA is funded by the STFC (UK), the ARC (Australia), the AAO, and the participating institutions. The GAMA website is \url{http://www.gama-survey.org/}.


\end{document}

%% file: table.tex
\begin{table}
\caption {Fits to the $M_{*}$-$j$ relations and to the $M_{*}$-$j$-$n_{r}$ and $M_{*}$-$j$-$\lambda_{R}$ planes. 
Scatters (rms) are orthogonal to the best fit.}
\[
\label{table}
\scriptsize
\begin{array}{cccccc}
\hline\hline
\noalign{\smallskip}
\multicolumn{6}{c}{\rm \log({\it j}/kpc~km~s^{-1})={\it a}\times \log({\it M_{*}}/M_{\odot})+{\it b}}\\
\noalign{\smallskip}
j_{*p} & a  & b  &  \multicolumn{2}{c}{rms} & N_{gal}\\
\noalign{\smallskip}
\hline
All          &  0.72\pm0.06   & -5.49\pm0.64   &  \multicolumn{2}{c}{0.27 } & 297 \\
E            &  0.83\pm0.18   & -7.24\pm1.98   &  \multicolumn{2}{c}{0.21}  & 26 \\ 
E/S0-S0      &  0.84\pm0.07   & -6.99\pm0.80   &  \multicolumn{2}{c}{0.17 } & 67 \\
S0/Sa-Sb     &  0.98\pm0.08   & -8.19\pm0.81   &  \multicolumn{2}{c}{0.16 } & 112\\
Sbc~or~later &  1.00\pm0.12   & -8.03\pm1.26   &  \multicolumn{2}{c}{0.18 } & 86 \\
\noalign{\smallskip}
j_{*} &          &                &         &    & \\
\noalign{\smallskip}
\hline
All           &  0.64\pm0.04   & -4.31\pm0.46   &   \multicolumn{2}{c}{0.22} &  297 \\
E             &  0.73\pm0.18   & -5.56\pm1.88   &   \multicolumn{2}{c}{0.21} &  26 \\ 
E/S0-S0       &  0.78\pm0.06   & -5.98\pm0.68   &   \multicolumn{2}{c}{0.15} &  67  \\
S0/Sa-Sb      &  0.96\pm0.07   & -7.58\pm0.73   &   \multicolumn{2}{c}{0.14} &  112  \\
Sbc~or~later  &  0.80\pm0.09   & -5.71\pm0.89   &   \multicolumn{2}{c}{0.14} &  86 \\
\noalign{\smallskip}
j_{p~gas} &          &                &         &     &    \\
\noalign{\smallskip}
\hline
All  &  0.68\pm0.03   & -4.75\pm0.27   &   \multicolumn{2}{c}{0.32} & 397 \\
\noalign{\smallskip}
j_{gas} &          &                &             &     \\
\noalign{\smallskip}
\hline
All  &  0.65\pm0.02   & -4.12\pm0.23   &   \multicolumn{2}{c}{0.28}  & 397 \\
\hline
\hline
\noalign{\smallskip}
\noalign{\smallskip}
\noalign{\smallskip}
\multicolumn{6}{c}{\rm \log({\it j}/kpc~km~s^{-1})={\it a}\times \log({\it M_{*}}/M_{\odot})+{\it b}\times \log({\it n}) + {\it c}}\\
\noalign{\smallskip}
           & a  &  b  &  c & rms & N_{gal} \\
\hline
j_{*p} & 1.22\pm0.07 & -1.86\pm0.13  & -10.09\pm0.73  & 0.13  & 297\\
j_{*}   & 1.05\pm0.06 & -1.38\pm0.10  & -8.18\pm0.56  & 0.12  & 297 \\
\hline
\hline
\noalign{\smallskip}
\noalign{\smallskip}
\noalign{\smallskip}
\multicolumn{6}{c}{\rm \log({\it j}/kpc~km~s^{-1})={\it a}\times \log({\it M_{*}}/M_{\odot})+{\it b}\times \log({\it \lambda_{R}}) + {\it c}}\\
\noalign{\smallskip}
           &  a  &  b  &  c & rms & N_{gal} \\
\hline
j_{*p} & 0.70\pm0.02 & 1.41\pm0.04  & -4.64\pm0.22  & 0.05 & 297\\
j_{*}   & 0.70\pm0.03 & 1.13\pm0.05  & -4.47\pm0.26  & 0.08 & 297  \\
\hline
\hline
\end{array}
\]
\end{table}